\documentclass[noshowkeys,amsmath,superscriptaddress,letterpaper,reprint,nofootinbib,aps]{revtex4-1}
\usepackage{graphicx}
\usepackage[lighttt]{lmodern}
\usepackage{amsfonts}
\usepackage{amssymb}
\usepackage{bm}
\usepackage{xcolor}
\usepackage[version=3]{mhchem}
\usepackage{csquotes} 
\usepackage{listings}
\usepackage{todonotes}
\usepackage{microtype}
\usepackage[linkcolor = blue, citecolor = blue, urlcolor = blue, colorlinks = true,pdfcreator=not,pdfproducer=latex]{hyperref}
\usepackage{epstopdf}

\newcommand{\es}{\textsf{ESPResSo}}
\newcommand{\esfour}{\es~4.0}
\DeclareMathOperator{\sgn}{sgn}

\usepackage{lmodern} 

\definecolor{stringblue}{rgb}{0.09,0.211,0.57}
\definecolor{codered}{rgb}{0.655,0.1137,0.3747}
\definecolor{deepgreen}{rgb}{0,0.5,0}
\definecolor{commentgray}{rgb}{0.4,0.4,0.4}
\definecolor{framegray}{rgb}{0.8,0.8,0.8}
\lstdefinestyle{pypressostyle}{
  language=Python,
  belowcaptionskip=1\baselineskip,
  breakindent=5pt,
  breaklines=true,
  xleftmargin=\parindent,
  showstringspaces=false,
  stringstyle=\color{orange},
  belowskip=\bigskipamount,
  aboveskip=\bigskipamount,
  basicstyle=\ttfamily\small,
  otherkeywords={self},             
  keywordstyle=\bfseries\color{codered},
  moredelim=*[s][]{.}{\ },
  moredelim=*[s][]{\ }{.},
  moredelim=*[s][]{.}{\[},
  emph={accuracy,
        actors,
        Actors,
        Actor,
        add,
        add_bond,
        add_reaction,
        Analysis,
        analyze,
        AngleCosine,
        AngleCossquare,
        AngleHarmonic,
        bjerrum_length,
        BondedInteraction,
        BondedInteractionNotDefined,
        BondedInteractions,
        bonded_inter,
        bonded_interaction_classes,
        box_l,
        cellsystem,
        CellSystem,
        constant_pH,
        ConstantpHEnsemble,
        COORDS_ALL_FIXED,
        COORDS_FIX_MASK,
        COORD_FIXED,
        cuda_init,
        CudaInitHandle,
        cutoff,
        Dihedral,
        default_charges,
        electrostatics,
        epsilon, 
        espressomd,
        exclusion_radius,
        features,
        FeneBond,
        FlowField,
        galilei,
        GalileiTransform,
        gamma,
        harmonic,
        HarmonicBond,
        HarmonicDumbbellBond,
        highlander,
        id,
        integrate,
        integrator,
        Integrator,
        interactions,
        lennard_jones,
        minimize_energy,
        MinimizeEnergy,
        non_bonded_inter, 
        NonBondedInteraction,
        NonBondedInteractionHandle,
        NonBondedInteractions,
        OifGlobalForces,
        OifLocalForces,
        OrderedDict,
        Overlapped,
        part,
        particle_data,
        PARTICLE_EXT_FORCE,
        PARTICLE_EXT_TORQUE,
        ParticleHandle,
        ParticleList,
        ParticleSlice,
        polymer,
        Polymer,
        pos,
        product_coefficients,
        product_types,
        P3M,
        q,
        reaction,
        reactant_coefficients,
        reaction_ensemble,
        reactant_types,
        ReactionEnsemble,
        RigidBond,
        run,
        set_params,
        set_steepest_descent,
        shift,
        sigma,
        steps,
        System,
        Tabulated,
        temperature,
        ThereCanOnlyBeOne,
        thermostat,
        Thermostat,
        time_step,
        update_wrapper,
        utils,
        v,
        Virtual},          
  emphstyle=\bfseries\color{deepgreen},    
  commentstyle=\color{commentgray}\ttfamily,
  stringstyle=\color{stringblue},
  frame=tb,                         
  rulecolor=\color{framegray},
  showstringspaces=false            %
}

\lstnewenvironment{pypresso}{\lstset{style=pypressostyle}}{}

\newcommand\pypressoinline[1]{{\lstset{style=pypressostyle}\lstinline!#1!}}

\lstdefinestyle{cxx}{
  language=C++,
  belowcaptionskip=1\baselineskip,
  breaklines=true,
  xleftmargin=\parindent,
  showstringspaces=false,
  stringstyle=\color{orange},
  belowskip=\bigskipamount,
  aboveskip=\bigskipamount,
  basicstyle=\ttfamily\small,
  otherkeywords={self},             
  keywordstyle=\bfseries\color{codered},
  moredelim=*[s][]{.}{\ },
  moredelim=*[s][]{\ }{.},
  moredelim=*[s][]{.}{\[},
  emphstyle=\bfseries\color{deepgreen},    
  commentstyle=\color{commentgray}\ttfamily,
  stringstyle=\color{stringblue},
  frame=tb,                         
  rulecolor=\color{framegray},
  showstringspaces=false            %
}

\lstnewenvironment{cxx}{\lstset{style=pypressostyle}}{}

\begin{document}

 
\title{\esfour{} -- An Extensible Software Package for Simulating Soft Matter Systems}

\author{Florian Weik}
\email{fweik@icp.uni-stuttgart.de}
\affiliation{Institut f\"ur Computerphysik, Universit\"at Stuttgart, Allmandring 3, 70569 Stuttgart, Germany}
\author{Rudolf Weeber}
  \affiliation{Institut f\"ur Computerphysik, Universit\"at Stuttgart,
    Allmandring 3, 70569 Stuttgart, Germany}
\author{Kai Szuttor}
  \affiliation{Institut f\"ur Computerphysik, Universit\"at Stuttgart, Allmandring 3, 70569 Stuttgart, Germany}
\author{Konrad Breitsprecher}
  \affiliation{Institut f\"ur Computerphysik, Universit\"at Stuttgart, Allmandring 3, 70569 Stuttgart, Germany}
\author{Joost de Graaf}
  \affiliation{Institute for Theoretical Physics, Center for Extreme Matter and Emergent Phenomena, Utrecht University, Princetonplein 5, 3584 CC Utrecht, The Netherlands}
\author{Michael Kuron}
  \affiliation{Institut f\"ur Computerphysik, Universit\"at Stuttgart, Allmandring 3, 70569 Stuttgart, Germany}
\author{Jonas Landsgesell}
  \affiliation{Institut f\"ur Computerphysik, Universit\"at Stuttgart, Allmandring 3, 70569 Stuttgart, Germany}
\author{Henri Menke}
  \affiliation{Institut f\"ur Computerphysik, Universit\"at Stuttgart, Allmandring 3, 70569 Stuttgart, Germany}
  \affiliation{Department of Physics, University of Otago, P.O. Box 56, Dunedin 9054, New Zealand}
\author{David Sean}
  \affiliation{Institut f\"ur Computerphysik, Universit\"at Stuttgart, Allmandring 3, 70569 Stuttgart, Germany}
\author{Christian Holm}%
\email{holm@icp.uni-stuttgart.de}
\affiliation{Institut f\"ur Computerphysik, Universit\"at Stuttgart, Allmandring 3, 70569 Stuttgart, Germany}

\date{\today}


\begin{abstract}
  \es{} is an extensible simulation package for research on soft
  matter. This versatile molecular dynamics program was originally
  developed for coarse-grained simulations of charged systems
  [Limbach~\textit{et al.}, Comput. Phys. Commun.~\textbf{174}, 704
  (2006)].  The scope of the software has since broadened
  considerably: \es{} can now be used to simulate systems with length
  scales spanning from the molecular to the colloidal.  Examples
  include, self-propelled particles in active matter, membranes in
  biological systems, and the aggregation of soot particles in process
  engineering.  \es{} also includes solvers for hydrodynamic and
  electrokinetic problems, both on the continuum and on the explicit
  particle level.  Since our last description of version 3.1
  [Arnold~\textit{et al.}, Meshfree Methods for Partial Differential
  Equations VI, Lect. Notes Comput. Sci. Eng.~\textbf{89}, 1 (2013)],
  the software has undergone considerable restructuring.  The biggest
  change is the replacement of the Tcl scripting interface with a much
  more powerful Python interface. In addition, many new simulation
  methods have been implemented.  In this article, we highlight the
  changes and improvements made to the interface and code, as well as
  the new simulation techniques that enable a user of \esfour{} to
  simulate physics that is at the forefront of soft matter research.
\end{abstract}

\maketitle

\section{\label{sec:intro}Introduction}

Soft matter~\cite{degennes92a, doi13a, barrat03a} is an established
field of research that concerns itself with systems ranging from the
nanometer scale up to tens of micrometers, existing on the interface between physics, chemistry, and biology.  These systems are classified as ``soft'' because the relevant energy scale of interactions is typically
comparable to the thermal energy, making them easy to deform through the application of external forces and fields.  The broadness of the field is exemplified by the subjects covered therein: polymer science~\cite{doi88a, rubinstein03a}, colloidal
science~\cite{verwey48a, lowen01a}, liquid
crystals~\cite{chandrasekhar92a, hamley03a}, biological
physics~\cite{nelson04a, levental07a}, food science~\cite{ubbink08a}, material science~\cite{polarz02a}, to name but a few.

Statistical techniques may be employed to describe soft matter systems, as thermal fluctuations and thus Brownian motion play an important role and allow the system to readily explore phase space. The field has thus proven itself to be a fruitful playground for shaping the community's understanding of statistical physics concepts~\cite{frey05a, frenkel02a}, as well as emergent non-equilibrium phenomena~\cite{seifert12a, cates12a, fodor18a}.  The
delicate interplay of energy and entropy that is found in many colloidal suspensions leads to a plethora of
complex structural properties that are difficult to grasp with pen and
paper.  Indeed, the soft matter community owes a great deal to
computer simulations for its current level of understanding. 

The use of computers in soft matter traces its root to first molecular-dynamics (MD) simulations of phase separation in a system of hard spheres~\cite{alder57a, alder62a}.  From these humble beginnings, computer simulation grew out to a new field that bridges experimental and theoretical efforts.  These methods
came to their own in soft matter, as the typical size and nature of the interactions found therein, lend themselves well to coarse graining. This is a process by which many of the microscopic freedoms that govern a system may be captured in terms of effective interactions on a larger scale.  The use of effective interactions enables the simulation of emergent effects
using simple models at a fraction of the computation effort that is
required to account for the full microscopic details.  This has played a decisive role in elucidating,~\textit{e.g.}, the
guiding principles for phase behavior of complex
liquids~\cite{pusey94a, bates98a, roji99a, leunissen05a}, the
structure of ferrofluids~\cite{camp00a, wang02a, klapp04a,
klinkigt13a}, the structure of polyelectrolytes~\cite{stevens93b,
dobrynin96a, micka99a, limbach02c}, and the swelling of hydrogels and
ferrogels~\cite{schneider02a, yan03a, mann04a, weeber12a, weeber18a}.

The MD simulation package \es{} is one of the most broad, efficient, and well-suited platforms for the study of coarse-grained models of soft matter. Only LAMMPS\footnote{\url{https://lammps.sandia.gov}}~\cite{plimpton95a}
and \textsf{ESPResSo++}\footnote{\url{http://www.espresso-pp.de/}} \cite{guzman18a-pre}
approach \es{} in terms of the scope of their simulation
methods and features. Here, we should emphasize that while \es{} is capable of simulating molecular fluids, its scope does not cover atomisticly resolved systems. There are a number of better-suited, open-source softwares
available for atomistic simulations, such as GROMACS\footnote{\url{http://www.gromacs.org}}\cite{berendsen95a, vanderspoel05a, pronk13a},
NAMD\footnote{\url{https://www.ks.uiuc.edu/Research/namd/}}\cite{nelson96a,
phillips05b, phillips14a}, and AMBER\footnote{\url{http://ambermd.org}}\cite{amber95a, amber05a}, to name just a few of the most commonly used ones.

The aim of this paper is to present the improvements made in
\esfour{} compared to our previous release \es~3.1~\cite{arnold13a}.  We have also taken the opportunity to introduce readers to our software package and encourage them to consider \es{}, both for use in soft matter research, and as a platform for
method and algorithm development.  A strong community is key to maintaining and extending a strong and reliable software for research.  We will provide examples of cutting-edge physics that can be simulated using \esfour{}, which will hopefully create interest in tyring out \es{} and potentially joining our thriving user and developer community.

In brief, this paper covers the following subjects.  We start by giving a broad introduction to \es{}, discussing its history, goals, and community in Section~\ref{sec:espresso}.  This will set the background to the developments made for our \esfour{} release.  Next, we focus on the new interface (Section~\ref{sec:python}), the interaction with external packages (Section~\ref{sec:integrate}), user interaction with the software in terms of visualization and output (Sections~\ref{sec:visualize} and~\ref{sec:output}, respectively), and changes to the new code development procedures and practices (Section~\ref{sec:engineer}) that we adopted with \esfour{}. Finally, we introduce the standout physical methods and models that were recently introduced within \es{}: energy minimization routines (Section~\ref{sec:minimize}), cluster analysis (Section~\ref{sec:cluster}), chemical reactions (Section~\ref{sec:reaction}), polarizable molecules (Section~\ref{sec:drude}), and self-propelled particles (Section~\ref{sec:active}). We end with a summary and outlook in Section~\ref{sec:conclude}.


\section{\label{sec:espresso}Extensible Simulation Package for
Research on Soft Matter Systems}

\subsection{\label{sub:espresso_intro}The Idea behind \es{}}

\es{} development began in 2001.  Its name is an acronym for
\textbf{E}xtensible \textbf{S}imulation \textbf{P}ackage for
\textbf{Res}earch on \textbf{So}ft matter systems, which encapsulates
the goal of the software project:  

On the one hand, the original development team wished
to create a highly parallel molecular dynamics program that could
deal efficiently with charged bead-spring models of polymers.  At that time, we
had learned to master electrostatic interactions in periodic
geometries~\cite{deserno98a, deserno98b} and had
also developed other fast electrostatic solvers for partially periodic
geometries~\cite{arnold02a, arnold02b, arnold02c, arnold02d,
arnold05a, arnold05b, arnold06a}. The resulting software saw its first use in the study of polyelectrolyte solutions and charged colloidal suspensions. 

On the other hand, the \es{} framework was designed to be \enquote{extensible}.  That is, its core architecture was created to facilitate the testing and integration of new algorithms. \es{} has seen many additions since 2001, living up to this extensible character~\cite{arnold13a}. This includes support for the simulation of hydrodynamic interactions via an efficient GPU implementation of a lattice Boltzmann solver, rigid-body dynamics, a scheme for the study of irreversible agglomeration, and
electrostatic solvers that allow for dielectric inclusions or even
locally varying dielectric permittivities~\cite{tyagi07a, tyagi08a,
tyagi10a, arnold13a, arnold13c, fahrenberger14a}. 

From the start, \es{} has a two-fold architecture, which roughly separates the developer from the user and promotes ease of use:  (i) Massively
parallel simulation routines were implemented in the C programming
language using the Message Passing Interface (MPI). These routines scale well up to hundreds of processors on large computing clusters. They also facilitate the addition of new algorithms, even by novice developers.  (ii) The simulation
core exposes a scripting interface to the user, such that they did not have to directly interact with the core code. This interface was based on the Tcl
language---the most suitable choice at that time to provide easy-to-use scripting. This gave the user programmatic control over the
simulation protocol, and enabled run-time analysis and visualization in a straightforward manner.

\subsection{\label{sub:espresso_develop}The Motivation for \esfour{}}

Every simulation package has to cope with changing computer architectures and programming models, ensure continued ease of development, and cater to users' expectations concerning interaction with the software. \esfour{} is our answer to the changes in the field of scientific computing that have taken place over the past five years. 

Today, the Python language has become the \textit{de-facto} standard
in scientific computing. There exists a vast ecosystem of third-party
Python packages for numerical algorithms, visualization, and
statistical analysis.  This motivated us to port the scripting
interface from Tcl to Python for the latest version of \es{}, as will be described in Section~\ref{sec:python}. We have also updated our online visualization to make the most of \esfour's new Python interface, see Section~\ref{sec:visualize}. Post-processing and visualization of large data sets is facilitated by the introduction of the HDF5 output format and parallellized output routines, see Section~\ref{sec:output}. Finally, the \es{} core has been converted to the C++ programming language, which has had added benefits both in terms of software development practices and in terms of performance, see Section~\ref{sec:engineer}.

To stay at the forefront of the field, it is not only necessary to
commit to the user and development experience, it is also critical to
keep up with the ever shifting interests of the user base. \es{} has
done so in the past~\cite{arnold13a} and continues this effort with
the \esfour{} release. Our simulation package now covers an even wider
range of physical systems of relevance to modern soft matter
research. This includes non-equilibrium phenomena in active matter,
which \esfour{} captures using model self-propelled
particles~\cite{fischer15a, degraaf15b, degraaf16a, degraaf16b}, see
Section~\ref{sec:active}. Going beyond soft matter, particle-based
methods have attracted interest from the engineering
community. Specifically, this community utilizes MD to model
large-scale physical processes such as soot aggregation~\cite{inci14a,
  inci17a} and air filtration~\cite{schober16a}, with the aim to
improve these processes. Section~\ref{sec:cluster} describes the
opportunities for these kinds of studies in our latest
release. However, as will become clear shortly, \esfour{} is capable
of much more.

\subsection{\label{sub:espresso_community}The User and Developer Community}

The \es{} package was set up as an open software project, committed to
open source development under the GNU General Public License (GPL), a
practice that we continue to adhere to.  Users of \es{} are supported
through mailing lists\footnote{\href{https://lists.nongnu.org/mailman/listinfo/espressomd-users
}{\nolinkurl{espressomd-users@nongnu.org}}} and
workshops, such as the annual \es{} summer schools, which have between
20 and 40 participants.  These workshops are often run as
CECAM\footnote{\url{https://www.cecam.org/}} tutorials, jointly funded
through grants from the German Research Foundation through the SFB~716
and the cluster of excellence SimTech (EXC~310).  The mailing lists
and workshops also allow the developer team to announce new features
implemented in the code base of \es{} to the global research
community.

\es{} is used by scientists all over the world, which is documented by
more than 431 citations on Web of Science since 2006 and 608 citations
on Google Scholar, as of this writing. Several research groups have
contributed algorithms to the core of \es{}, which include:
(i)~Continuum flow solvers (including ones on GPUs) and various methods to couple
them to MD~\cite{rohm12a, cimrak12a, cimrak14a, cimrak13a,
rempfer16a, guckenberger16a, baecher17a}.
(ii)~Advanced algorithms for rare-event sampling and the
reconstruction of free-energy landscapes~\cite{kratzer13a,
kratzer14a}.
(iii)~Monte-Carlo methods~\cite{samin13a}.
(iv)~An interface to Python analysis package MDAnalysis.
(v)~Extensions to the support for anisotropic particles.
(vi)~Methods to calculate dipolar interactions on the GPU.
Significant effort has been made to accommodate the various sources
of these contributions and ensure code quality of \esfour{}. This
includes putting in place an extended test infrastructure and
increasing the test coverage. Furthermore, all changes to the
software, including those made by the core team, undergo peer review
before they are merged, see Section~\ref{sec:engineer}.  

The latest version of \es{} and its
documentation can be found on the website
\url{http://espressomd.org}.  Ongoing development is organized
through the social code hosting platform
GitHub\footnote{\url{https://github.com/espressomd/espresso/}}, which
facilitates contribution handling through its collaborative features.

\section{\label{sec:python}The Python interface}

In recent years, Python has become increasingly popular in the
scientific community as a scripting language, in areas such as machine
learning (PyTorch, scikit-learn, Keras)~\cite{paszke17a, pedregosa11a,
  chollet17a}, symbolic mathematics (SymPy, SAGE)~\cite{meurer17a,
  stein05a}, data visualization (Matplotlib,
Mayavi)~\cite{hunter07a, ramachandran11a}, and numerical mathematics
(NumPy, SciPy, pandas)~\cite{jones01a, mckinney10a}.  In \esfour{}, we
have therefore decided to switch the interface scripting language from
Tcl to Python.  This allows for the quick generation of clear and
concise scripts that can harness the power of third party numerical
tools.  The \es{} core functionality can be accessed like any other
Python module by importing the
\lstinline[style=pypressostyle]|espressomd| module.  The whole
architecture is built around the
\lstinline[style=pypressostyle]|system| object which encapsulates the
physical state of the high-performance simulation core.

\es{} is a particle-based molecular dynamics simulator, \textit{i.e.},
all fundamental operations in the simulation setup revolve around
particles. Inspired by Python's list datatype and the NumPy
package~\cite{jones01a}, we introduced slicing operations on the list
of particles.  Many functions also accept lists and automatically
broadcast themselves to the individual list elements.  This often
reduces the number of lines of code significantly, \textit{e.g.},
placing several particles at random positions in the simulation box
can now be done in a single line:
\begin{lstlisting}[style=pypressostyle]
import numpy as np

system.part.add(pos=np.random.random((50, 3)) * system.box_l)

print(system.part[:].pos)
\end{lstlisting}

In Python, functions can be passed to other functions as
arguments. \esfour{} makes use of this, \textit{e.g.,} for particle
selection using a user-specified criterion:
\begin{lstlisting}[style=pypressostyle]
charged_particles = system.part.select(lambda p: p.q != 0.0)
\end{lstlisting}

Because the NumPy package~\cite{jones01a} is very popular in numerical
computing, including in undergraduate courses, we designed the
interface for maximal compatibility.  This allows the user to directly
pass on values returned from \es{} functions to a variety of NumPy
routines, \textit{e.g.}, to compute the mean, standard deviation, or
histograms.

We further make use of the object-oriented programming capabilities of
Python.  Instead of commands having an immediate effect, the user
instantiates objects of a class and adds these objects to the system
instance.  For example, when the user creates an object for the 
electrostatic solver, it is not immediately active, but has to be 
added to the list of actors first. This new design aims to reduce 
dependencies between different parts of the user-provided simulation
script and eliminate issues related to the order of commands.

\section{\label{sec:integrate}Integration with External packages}

\subsection{Using \es{} with other Python packages}

As we already mentioned earlier, the \es{} Python interface was
designed with NumPy in mind and most functions return datatypes which
are compatible with the routines offered by NumPy.  These routines are
well-maintained and supported by a large community with numerous online
resources and introductory material. 
One example are various statistics routines.
These include mean, variance and median calculations, but also \lstinline[style=pypressostyle]{numpy.histogram()}. As an example, a probability distribution of inter-particle distances can be used to obtain effective potentials via Boltzmann inversion \cite{reith03a}.
Furthermore, the confidence interval estimators from \lstinline[style=pypressostyle]{scipy.stats.t.interval()} can be used to control
the amount of sampling in a simulation.
Random number generation from \lstinline[style=pypressostyle]{numpy.random} and
  \lstinline[style=pypressostyle]{scipy.stats} can be used, e.g., for setting up a
randomized starting configuration.
Linear algebra methods from \lstinline[style=pypressostyle]{numpy.linalg} allow for easily calculating
the principal axes of and moments of inertia of a set of particles.
Lastly, new interpolated bonded and pair potentials in \es{} can be added by providing NumPy arrays for the energies and forces. 

It is also possible to create Jupyter notebooks \cite{kluyver16a} using \es{}.  These
interactive worksheets are popular for teaching and provide a great
way to nicely present results and the generating simulation code
alongside.  For example the introductory tutorials for \esfour{} are presented in
this fashion.

\esfour{} provides an interface to the MDAnalysis Python package,
which provides a large number of analysis routines for particle-based
data.  This includes the calculation of radial distribution functions,
density profiles, and the persistence length of polymer chains.

\subsection{Algorithms for Coulomb and dipolar interactions from the ScaFaCoS library}

The treatment of Coulomb and dipolar interactions has always been one
of the strong points of \es{}. Electrostatic interactions are
inherently important in many soft matter systems: \textit{e.g.}
colloidal suspensions are often stabilized by means of electrostatic
repulsion, the structure of polyelectrolytes depends on the
concentration of charged salt atoms, and electric fields are used for
the analysis and separation of particles in electrophoresis
applications.

Being long-ranged, electrostatic interactions have to be treated by
specialized algorithms in systems with periodic and mixed boundary
conditions.  \es{} already contains implementations of the
particle-particle-particle-mesh (P$^3$M) algorithm \cite{hockney88a} for fully periodic
systems, as well as the electrostatic layer correction (ELC) \cite{arnold02c} and MMM2D \cite{arnold02a}
scheme for 2D-periodic slit-pore geometries.  However, more algorithms
have been developed, designed for particular applications. Many of
these are available in the ScaFaCoS library~\cite{arnold13b}
(\textbf{Sca}lable \textbf{Fa}st \textbf{Co}ulomb
\textbf{S}olvers).  \esfour{} provides an interface to this library,
exposing all provided electrostatic solvers.  Moreover, an extension
to the ScaFaCoS library adds solvers for dipolar interactions with an
$\mathcal{O}(N \log N)$ scaling for periodic, mixed, and open boundary
conditions. With \esfour{}, this version of the library can be used
as a solver for magnetostatic interactions~\cite{nestler16a,
weeber18c-pre}.
The dipolar P$^2$NFFT solver for open boundaries is of particular relevance to research in magnetic soft matter: properties of such material can depend on the sample shape due to the demagnetization field \cite{raikher03a,morozov09a}.

\section{\label{sec:visualize}Online Visualization}

\begin{figure}[tb]
    \centering 
    \includegraphics[width=\columnwidth]{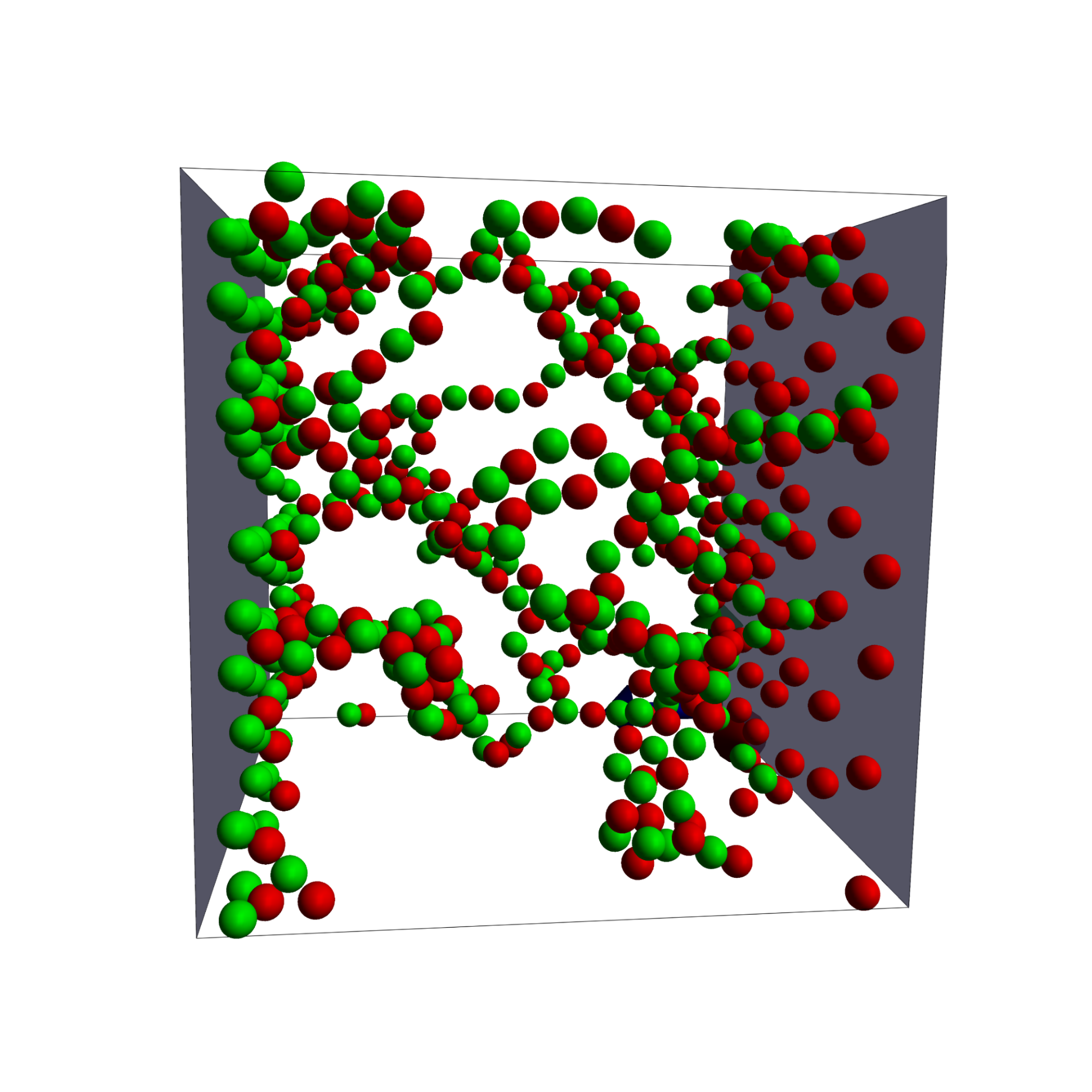} 
    \caption{Example of the OpenGL visualizer of \es{} for charged
      particles in a plate capacitor.  Positive (red) and negative
      (green) ions experience a constant-potential boundary
      condition.}
    \label{fig:visualizer_mmm2d} 
\end{figure}

\begin{figure}[tb] 
    \centering 
    \includegraphics[width=\columnwidth]{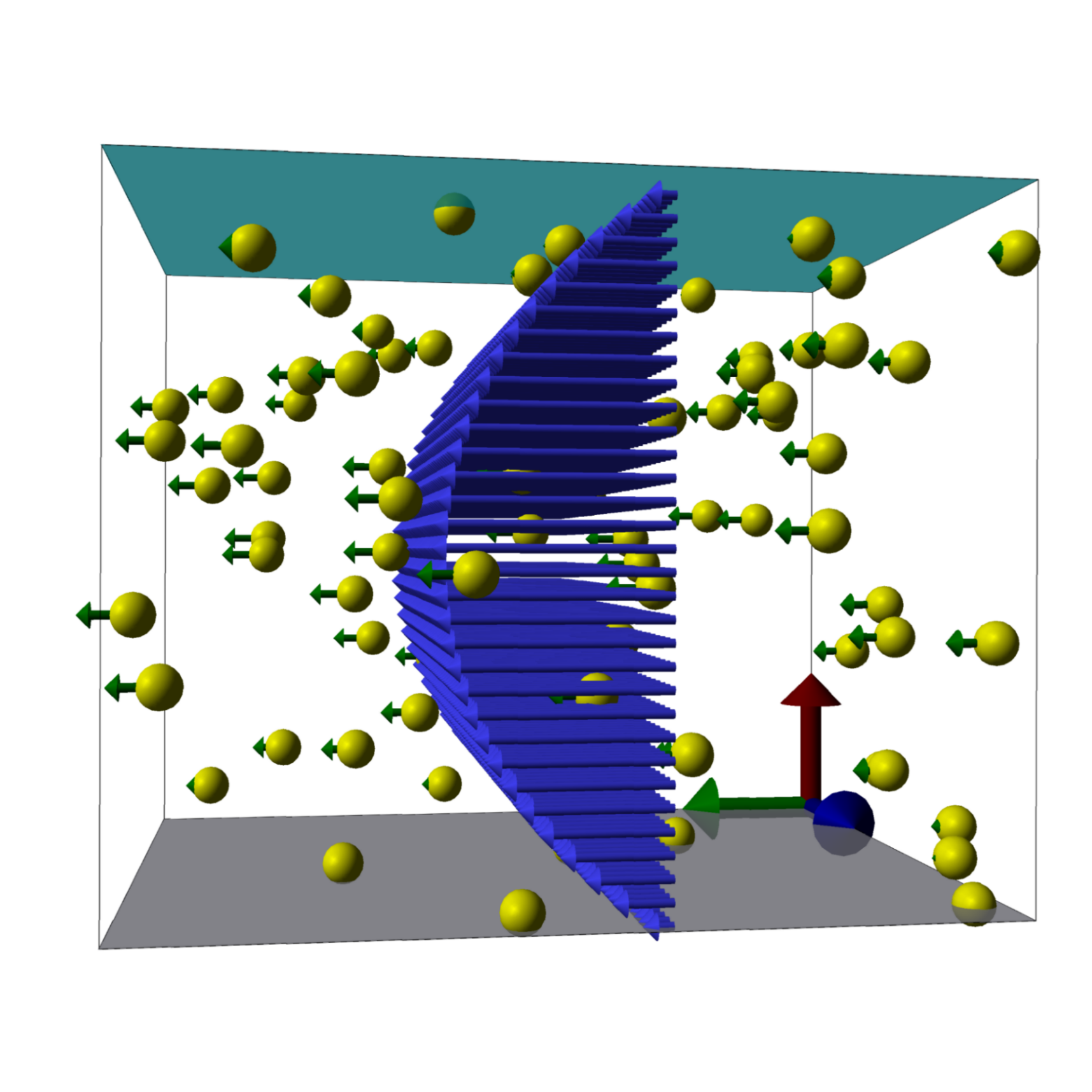} 
    \caption{Example of the OpenGL visualizer of \es{} for particles
      advected in a Poiseuille flow between two walls (Green arrows:
      particle velocity; blue arrows: fluid velocity).}
    \label{fig:visualizer_lb} 
\end{figure}

A graphical representation of the system is of great value when it comes to
presenting results of a simulation study. Also, it is common practice to
visualize intermediate results while setting up the simulation. This saves time, as
possible pitfalls related to positions, directions or orientations can be
rectified immediately when discovered via visualization. Usually, this is done by
writing out trajectories or volumetric data and utilizing external tools like
Paraview~\cite{ayachit15a} or VMD~\cite{humphrey96a}.  \es{} with
Python is well suited for this kind of workflow. Furthermore, \es{} features
two options for visualizing simulations while they are running, \textit{i.e.},
live or on-line visualization. The first one uses
Mayavi~\cite{ramachandran11a}, a Python framework for \enquote{3D scientific
data visualization and plotting} to visualize particles.  Mayavi has a
user-friendly graphical interface to manipulate the visual appearance and to
interact with the resulting representation.  Second, {\esfour} comes with a 3D
rendering engine that uses PyOpenGL~\cite{fletcher05a, shreiner99a}.  Both
visualizers are not primarily intended to produce print-quality renderings, but rather to
give a quick impression of the system setup and the equilibration process. Two
exemplary snapshots are shown in Figs.~\ref{fig:visualizer_mmm2d}
and~\ref{fig:visualizer_lb}; the corresponding Python scripts are available in
the distribution package of \esfour{}.

Our OpenGL visualizer has been crafted specifically for use with \es{}
and is capable of not only visualizing particles, but also several specific
features like constraints, detailed particle properties, the cell system, the
domain decomposition across processors, or fluid flows computed with the
lattice Boltzmann method~\cite{roehm11a} (Fig.~\ref{fig:visualizer_mmm2d}). It
has a large number of options to adjust colors, materials, lighting, camera
perspective, and many more. Vectorial particle properties can be visualized by
3D arrows on the particles to display forces, velocities, etc.  The
OpenGL visualizer can also be used for offline visualization to 
display completed simulations.  

A unique feature is the ability to interact with the running
simulation: Particles can be queried for their properties in real-time
by simply clicking on them in the 3D window. Most of the system
information like active interactions, actors and global properties can
be viewed in the visualization. In the Python script, the user can
assign callback functions to keyboard input, so that all exposed
system and particle properties (\textit{e.g.}, the temperature of the Langevin
thermostat or the external force on a particle group) can be
manipulated in real-time for testing and demonstration purposes. This
allows to use \es{} also as an educational tool for a broad range of
skill levels from basic physics to advanced simulation methods.

\section{\label{sec:output}Parallel Output}

Writing trajectories of a large set of particles to disk can be a bottleneck of
parallel MD simulations. For serial I/O the costs of writing the trajectory
include the communication to gather the data on the node that actually performs
the write. A second aspect of the performance cost is the amount of data that has
to be written to disk. Thus, in \esfour{} we added the possibility to write binary
files in parallel using the H5MD specification~\cite{buyl14a} which utilizes the HDF5 file
format~\cite{misc-hdf5}. H5MD files are intended to be self-contained and therefore
enhance the portability of simulation output. This more readily allows one to
use a wide class of analysis programs. The implementation is based on a wrapper
around the HDF5 library. In typical \es{} simulations, the I/O performance is
improved by using parallel output for particle numbers greater than
approximately $10^{4}$. Writing in this binary format is significantly faster
than more traditional ASCII text. The H5MD files written by \esfour{} may
conveniently be read into Python using the H5py module~\cite{collette14a}.
Simulation trajectories may also be visualized in VMD~\cite{humphrey96a} by using
its H5MD plugin. The H5MD output file contains a field that holds the source
script that was used for its creation. Thus the origin of the raw simulation
data can always be traced back to the \es{} Python script, which helps tremendously in
reproducing data.

For convenience, we still support \es{}'s VTF format which outputs system and
particle properties as ASCII text. However, the use of Python as a scripting
language opens the door to a large number of alternatives. Many aspects of a
simulation, as well as simulation parameters can, for instance, be stored in a
structured format using Python's \verb|pickle| module.
 
\section{\label{sec:engineer}Software Engineering and Testing}

As outlined in the previous section, the scripting interface is
powered by a high-performance simulation core.  The core of previous
versions of \es{} was implemented in the C programming language.  It
was only natural to also upgrade the core in a manner that mirrors the
new object-oriented approach of the interface.  Therefore, \esfour{}
has switched from C to C++. C++ is a modern object-oriented
programming language that allows for a more compact, readable code
style.  Just like Python, C++ benefits from an active community and a
rich ecosystem of third-party support libraries, \textit{e.g.}, the
Boost libraries~\cite{misc-boost}.  Thus, functionality that was
previously implemented in the simulation core can now be provided
externally. This reduces the code complexity and consequently the
maintenance effort.

\es{} is designed with high-performance computing in mind.  On
high-performance compute clusters, the environment is often times very
heterogeneous.  For example, external libraries provided by different
operating systems can vary noticeably, and only very recent C++
compilers fully support modern standards like C++11.  To ensure that
\es{} can be built and run on a large variety of different setups, it
turned out to be necessary to comprehensively test it on different
combinations of operating systems and compiler versions. \es{} has
always included a suite of test cases that could automatically run
small simulations to formally verify the physical correctness of the
implementation by comparing to established results. In \esfour{}, not
only have these tests been extended to cover more code, but also to
specifically test the correctness of certain core aspects, a method
called unit testing.

\subsection{Improvements in the simulation core}
\label{subsec:core_improvements}
The \es{} code base has been developed continuously for more than 15
years by a host of contributors of various backgrounds and
styles.  Thanks to the substantial overlap of language features
between C and C++, it was possible to switch the language with minimal
effort.  Even though C and C++ share a common heritage, the
programming paradigms are radically different and we are gradually
transforming the old C code into more modern C++ code.

New language and library features of C++, especially ones which became
available with the C++11 standard, are employed for more concise and
expressive design of our software. \es{} is continuously used in
day-to-day research activities and mainly developed by interested
domain researchers; hence a gradual modernization strategy was
necessary.  Among the many algorithms whose code has been refactored,
one central method is the traversal of particle pairs used for the
evaluation of short-range interactions. In \esfour{}, the traversal
has been separated from the kernel that evaluates the interactions for
each particle pair. Repetition of the complicated routine for the pair
energy, virial, and force calculation in the three types of particle
cell systems supported by \es{} is hence avoided. This has been
achieved by replacing several loops over the particles by a single
function that takes the kernels as template parameters. This allows us
to test key routines in isolation, independent of any interaction
potentials. Added benefits to this decoupling are, improved
performance and the ease by which new methods may be introduced by
(future) collaborators.

\subsection{Testing}

\es{} contains a test suite to verify the correctness of the code.
Hitherto, these tests were run by the maintainers before accepting code
contributions, but features without tests were not covered.  A formal
measurement of the code coverage was introduced because it proved
hard to detect which features are untested.  In preparation for this
release, we systematically improved coverage of previously untested
areas of code.  It is our experience that adding tests
to existing code not only ensures correctness, but also improves the
user experience by streamlining interfaces and avoiding
inconsistencies and confusion.

For example, \es{}'s GPU and CPU lattice Boltzmann (LB)
implementations had few systematic tests.  However, this is an
important method for many users of \es{}, and central for the
correctness of many soft matter simulations~\cite{duenweg09a}.  We
therefore introduced multiple new tests in this area that compare the
flow fields computed by the LB method, as well as its coupling to
particles, against stationary and time-dependent analytical solutions
and reference data.  These tests include direct verification of
momentum and mass conservation, as well as validation of the
statistical properties of the fluctuating variant of the LB and
particle coupling. This helped us to improve the existing code as well
as its interface.

In addition to the improved integration tests, this release includes
some unit tests.  Unit tests help identify issues on the level of
individual functions which make up the algorithms and give more
precise information where an issue occurred.  This further allows
changing the building blocks of an algorithm or their reuse across
different simulation methods.

\subsection{Continuous Integration}

As mentioned earlier, the development of \es{} is organized through
GitHub.  Contributors can submit
patches to fix a bug or introduce new features by means of a pull
request.  Whenever such a pull request is filed, the aforementioned
set of unit and integration tests is automatically run and the outcome
is communicated to GitHub where it is displayed in an informative
fashion.  This method of automatically running the test suite for
changes is called continuous integration.  Many online services exist
which are free to use for open source projects like \es{}.  Unfortunately, their free
service plans are too limited for the vast variety of setups we need
to test.  Therefore we mirror the \es{} repository to a private
instance of GitLab\footnote{GitLab is a free, self-hosted platform
  similar to GitHub, see \url{https://gitlab.com}.}.  This
instance manages a fleet of dedicated in-house computers and
distributes jobs for each configuration to test.  On top of the
automated testing, other developers are tasked with peer review of the
patches to ensure a consistent integration into the code base.

We also try to cover as many combinations of operating systems,
compilers, Python versions, and with and without GPU processors as
possible.  Our use of the Clang compiler~\cite{misc-clang} is
particularly noteworthy as it is able to perform static code analysis
to detect a large number of common programming errors and it is able
to generate code to automatically recognize undefined behavior during
runtime, thus often detecting bugs that are hard to discover for a
human software developer.

\section{External Fields}

\begin{figure}[tb]
    \centering 
    \includegraphics[width=.8\columnwidth]{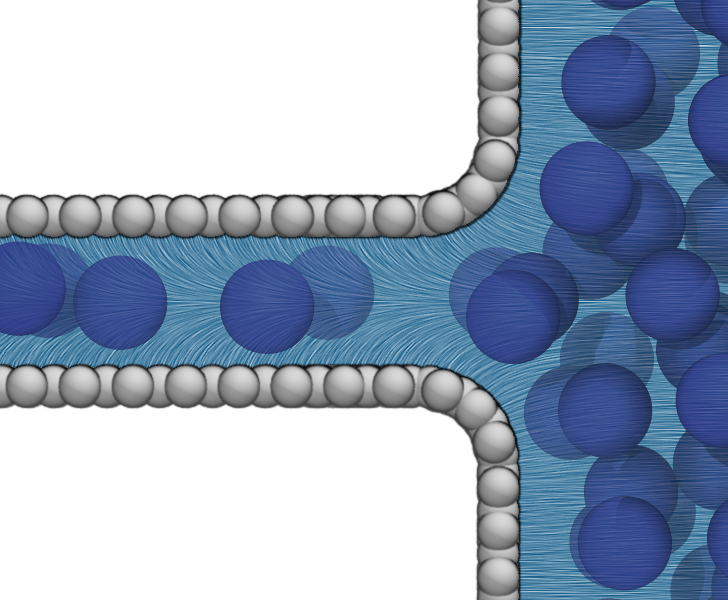} 
    \caption{Electric Field lines in a slit pore filled with an ionic liquid.}
    \label{fig:slit_pore} 
\end{figure}

\esfour{} implements a new extensible framework for coupling particles
to external fields, \textit{i.e.}, fields that are not caused by
particle interactions. 
An ``external field`` is composed of a field source and a coupling, which can be combined as needed.
An important use case for this feature are particles advected in a prescribed flow field under the assumption that they do not back-couple to that flow field. An example are soot particles on the nanometer scale in combustion exhausts \cite{inci17a,smiljanic18a-pre}.
Here the field source would be a flow field specified on a regular grid which is interpolated to the particle positions. The coupling term would be a viscous friction, combining the velocity difference between the particle and the flow field and a friction constant to a force.
Further use cases for external fields are complex boundary shapes as they are
found in super-capacitor electrodes as for example described by
\citet{breitsprecher18a}. In this application, electrodes with non-trivial
geometry are considered. The electric field caused by the electrodes is
pre-computed by solving Poisson's equation for the system, and the resulting
field is then superimposed on the inter-particles forces in the MD simulations
as an interpolated external electric field. A snapshot of particles and electric
field lines in such a system is shown in Fig.~\ref{fig:slit_pore}.
The feature can also be used for Lennard-Jones particles in front of convex surfaces. Such a particle can interact with an extended part of the surface. This cannot be described by \es{}'s constraint feature, which only accounts for an interaction along the surface normal vector.

Let us now look at the two components of external fields in detail.
The field source is described as a global scalar or
vectorial field.
Either, user-provided data is interpolated from a
regular grid onto the particle positions, a constant value throughout the full domain is used, or an affine map $A$ according to
\begin{equation}
  \bm{u}(\bm{r}) = A \bm{r} + \bm{b},
\end{equation}
with a user-provided matrix $A$ and shift $b$ is evaluated.

The coupling maps a particle
property and the value of the field source at the particle position to
a force or potential.  In the latter case the force is calculated by
invoking the coupling with the gradient of the field.  For example a
gravitational field can be defined as a constant scalar field source
coupling to the particle mass.

Presently there are several possibilities for the particle coupling,
including but not limited to a particle's scalar mass or charge, or
the particle's velocity vector $\bm{v}$ using a viscous coupling,
\textit{e.g.},
\begin{equation}
\label{eqn:viscous}
  \bm{F}_i = -\gamma (\bm{v}_i - \bm{u}).
\end{equation}
As a simple example we consider a two-dimensional system comprising
particles with excluded-volume interactions in a square simulation box
of side length $2\pi$.  We combine interpolated flow field data with a
viscous coupling.  The flow field is a discretized static Taylor-Green
vortex, specified by
\begin{equation}
  \label{eq:taylor_green_vortex}
  \begin{split}
    u_x &= \cos(x) \sin(y) ,\\
    u_y &= -\sin(x) \cos(y),
  \end{split}
\end{equation}
where $\bm{u}$ is the flow velocity.
It has the vorticity
\begin{equation}
  \label{eq:taylor_green_vorticity}
  \omega \equiv \nabla \times \bm{u} =  2 \left| \cos(x) \cos(y) \right| .
\end{equation}

\begin{figure}[tb]
    \centering 
    \includegraphics[width=.8\columnwidth]{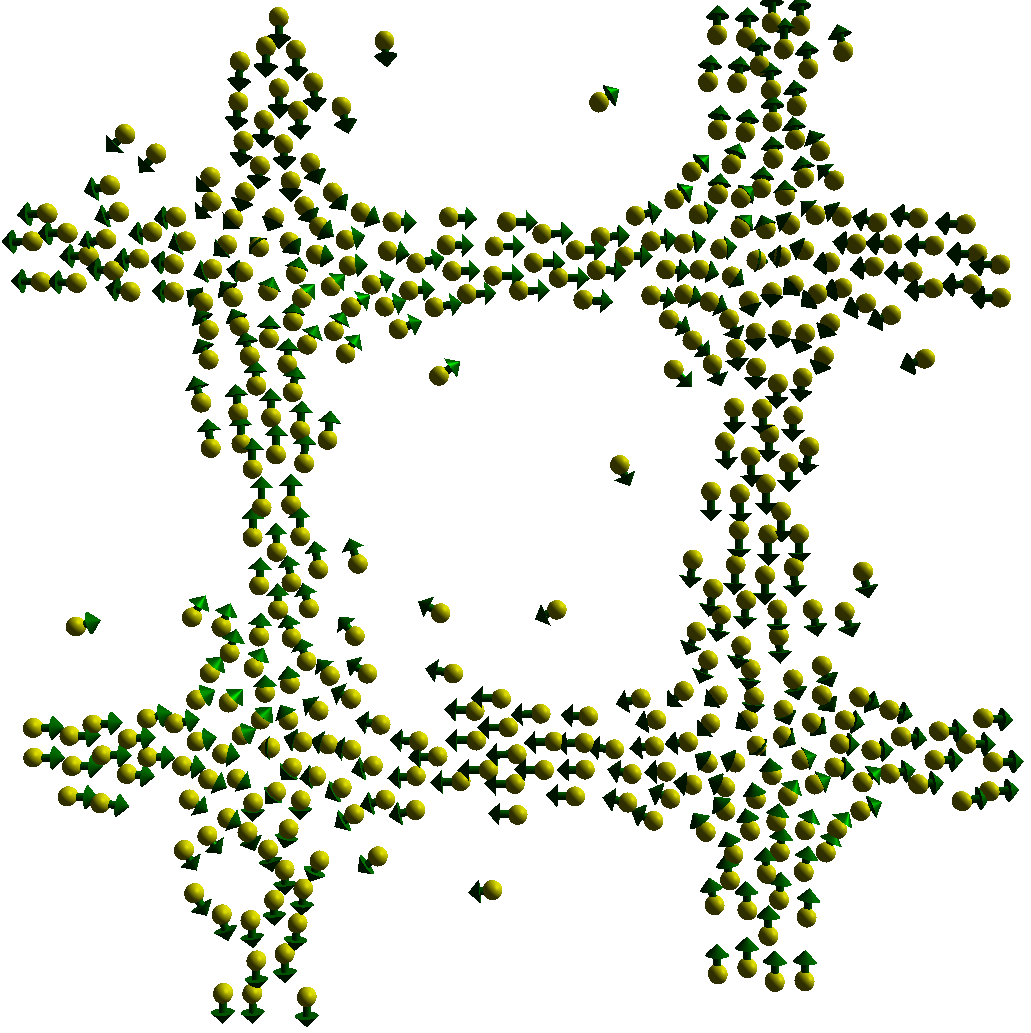} 
    \caption{Example for the use of external fields in \esfour{}.  We
      show the stationary state of particles in a static Taylor-Green
      vortex flow.  The green arrows denote the particle velocity and
      direction.}
    \label{fig:vortex} 
\end{figure}

In Fig.~\ref{fig:vortex} we show a snapshot of a simulation where the
particles were initially placed at random positions distributed
equally over the simulation volume. They are driven out of the areas
with high vorticity $\omega$ by inertial forces and accumulate
elsewhere.  A
complete simulation script can be found in the supplementary
information.

\section{\label{sec:minimize}Energy Minimization}

Since bulk molecular dynamics are typically set up with random
particle positions, initial configurations often have a lot of overlap
between neighboring particles.  This leads to problems with numerical
stability due to very high energies. To remove this overlap there are
two common methods. Limiting the forces between particles to a maximum
value and performing an energy minimization step after the initial
setup. While it was possible to cap the the forces in previous version
of \es{}, now also energy minimization by the steepest descent method
is available. It allows relaxing the position as well as the
orientation of the particles by running the following iterative
scheme:
\begin{align}
  \label{eq:displacement_rule}
  \Delta_i^{x,y,z} &= \sgn(F_i^{x,y,z}) \min(\gamma F_i^{x,y,z}, \Delta_{\text{max}} ) \\
  x_i^{x,y,z} &= x_i^{x,y,z} + \Delta_i^{x,y,z} ,
\end{align}
while $\max_i{|F_i|} \geq F_{\text{max}}$.  Here the $x_i^{x,y,z}$ and
$F_i^{x,y,z}$ are the components of the position and force of the
$i$th particle. That is the update rule and limits are applied by
component. $\gamma$ and $F_{\text{max}}$ are relaxation parameters
provided by the user. The orientation of the particle is relaxed in a
similar fashion.  Steepest descent energy minimization is available as
an alternative to the integrator in \es{}.

\begin{figure}[tb]
  \begin{tabular}{@{}ccc@{}}
  \includegraphics[width=0.3\columnwidth]{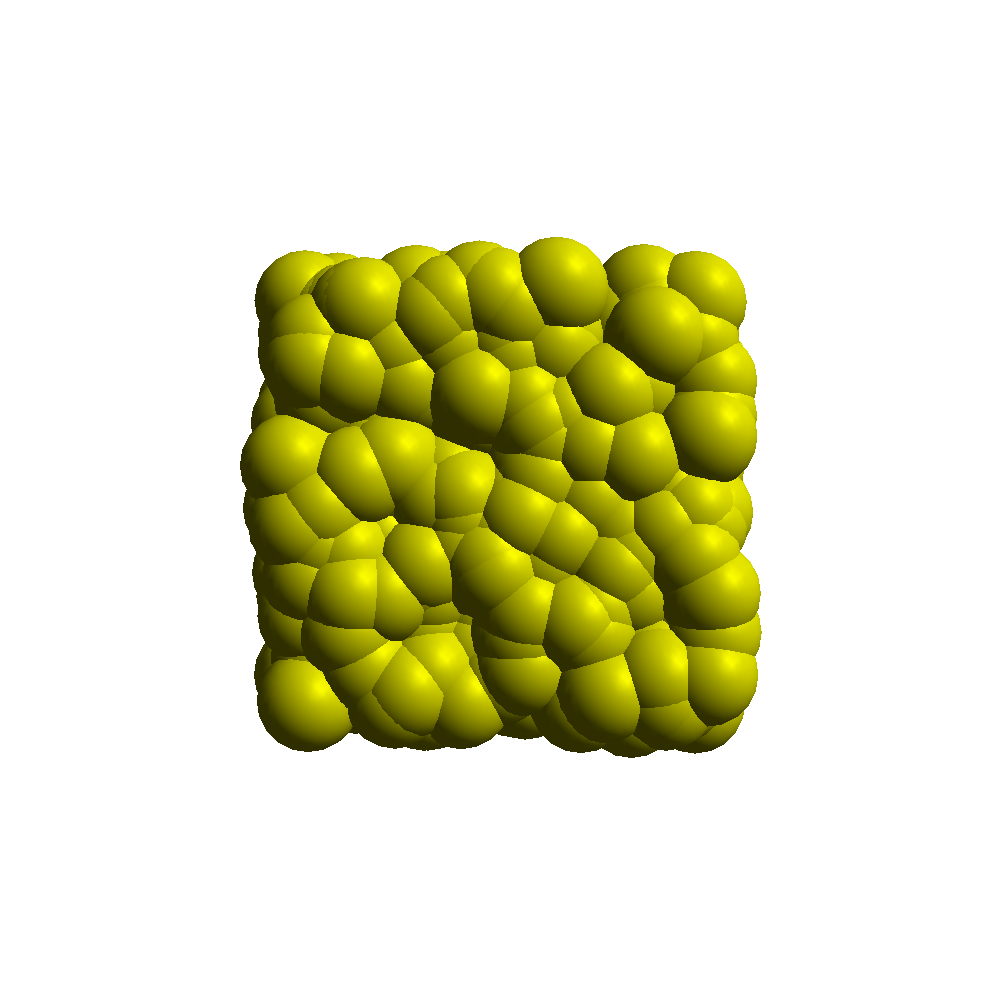}
    & \includegraphics[width=0.3\columnwidth]{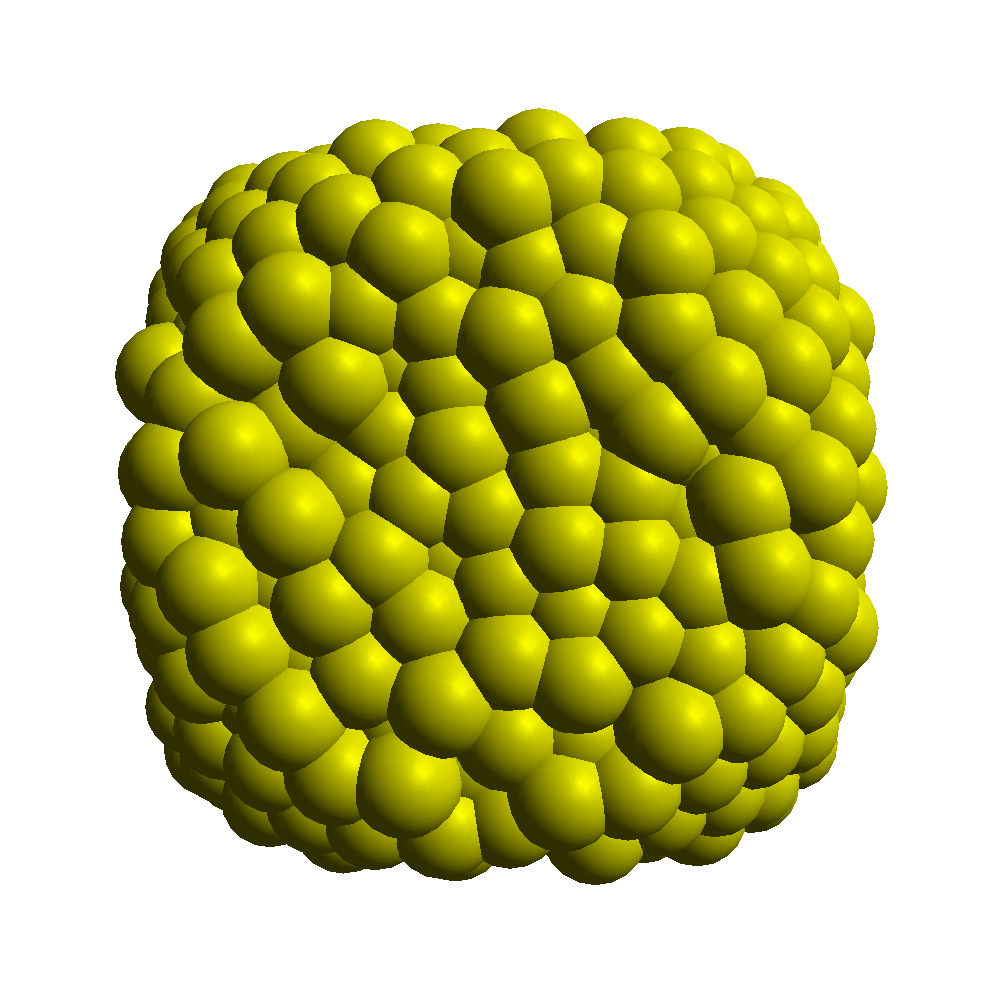}
    &  \includegraphics[width=0.3\columnwidth]{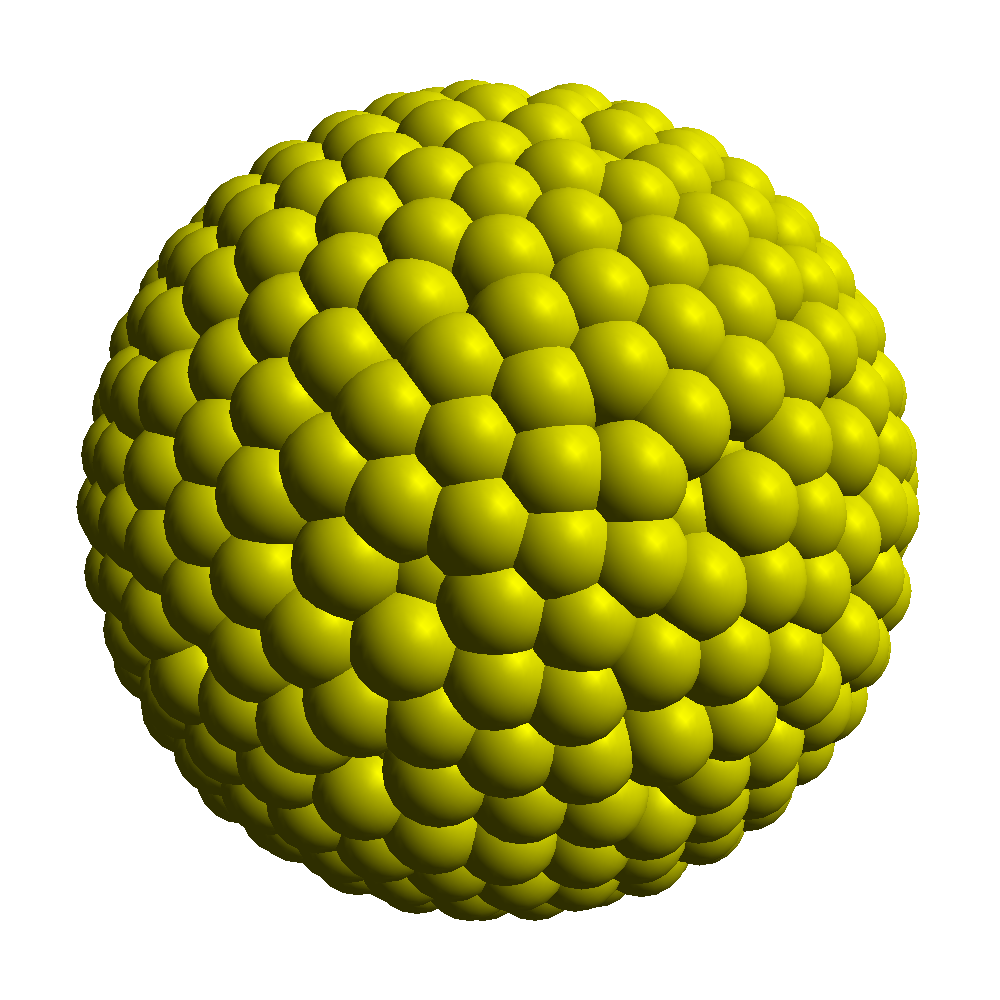} \\
    (a) & (b) & (c) \\
  \end{tabular}
  \caption{(a) Initial configuration of the particles filling the
    raspberry. (b) Configuration after 100 steps energy relaxation. (c)
    Configuration after 1000 steps.}
  \label{fig:filled_raspberry}   
\end{figure}

In soft matter simulations the solvent is often treated
on a coarse-grained level using a lattice Boltzmann fluid, which
interacts with the particles of the system via so-called point
coupling~\cite{duenweg09a}. This type of coupling limits the maximal
coupling strength for one particle which can lead to unrealistic
transport properties for larger objects like colloids. One way to
overcome this problem is to use so-called raspberry models \cite{lobaskin04a}, rigid
bodies of multiple particles, which couple to the fluid in multiple
points to reduce interpolation artifacts and can achieve higher
friction with the fluid. This can help to obtain better transport
coefficients~\cite{fischer15a, degraaf15b}.  To setup such a raspberry
particle, its volume is filled with particles in a homogeneous manner. A
common way to do this is to put the particles into a spherical
constraint, give them a repulsive interaction and relax the energy.
This can now be easily implemented in \es{}. As an example,
Fig.~\ref{fig:filled_raspberry} show the initial and relaxed
configuration of 1400 particles in a spherical constraint with a
soft-core Weeks-Chandler-Andersen interaction. Initially the particles
are placed randomly in a cube contained by the sphere, then the energy
minimization is performed. For the simulation script and detailed
parameters please refer to the supplementary information.

\section{\label{sec:cluster}Cluster Analysis}

An analysis of the spatial clustering of particles is an important
tool in many fields of soft matter science.  Some examples include the
nucleation of crystals~\cite{radu13a,kratzer15a}, clustering of
magnetic nanoparticles in magnetic soft matter
systems~\cite{butter03a, cerda08a, weeber13a, donaldson15a}, and the
irreversible agglomeration of soot particles in combustion
processes~\cite{attili14a, inci14a, inci17a}.  Clusters are typically
defined by means of a criterion for a given pair of particles which
takes the form of an equivalence relation. That is, if particles $a$
and $b$ are \enquote{neighbors} and particles $b$ and $c$ are
\enquote{neighbors}, all three particles belong to a cluster.  The
pair criterion is chosen based on the application. For crystallization
studies, a local bond order parameter~\cite{lechner08a} above a
certain threshold on both particles is a common choice. For magnetic
systems, a combination of distance and either dipole configuration or
pair energies are used~\cite{cerda08a, weeber13a}. For the
agglomeration of soot particles, the bonds created by the collision
algorithms are used~\cite{inci17a}.

While it is possible to perform cluster analysis in the
post-processing state of a simulation, this requires the storage of a
significant number of snapshots of the simulation. Due to the 
potentially large amount of storage space this might take, as well as the extra time it takes to write the snapshot to disk,
this is often not a good approach.  \esfour{}
therefore introduces an online cluster analysis which can be run from
within the simulation. The implementation is loosely based on the
Hoshen-Kopelman scheme~\cite{weeks71a}, but does not use a lattice.
The procedure is as follows: All pairs of particles are examined. If
they are \enquote{neighbors} as determined by a user-specified pair
criterion, there are several possible cases.
\begin{itemize}
\item If neither particle belongs to a cluster, a new clusters is
  created and both particles are marked as members.
\item If one particle is part of a cluster and the other is not, the
  free particle is added to the cluster.
\item If the two particles belong to different clusters, a note is
  made that the two clusters are one and the same and need to be
  merged later.
\end{itemize}
The clusters are labeled by numeric IDs, assigned in ascending order.
After all pairs of particles are examined, the clusters marked as
identical in the previous step are merged. Cluster IDs are traversed
in descending order and merged clusters receive the lower of the two
cluster IDs. In this way, the merging can be done in a single pass.
\esfour{} currently contains pair criteria based on inter-particle
distance, pairwise short-range energy, and the presence of a bonded
interaction. Further criteria can be added easily.  The cluster
analysis is organized around a \verb|ClusterStructure| object, which
provided the methods to run analysis as well as access to the clusters
found. Furthermore, some analysis routines are provided on a
per-cluster level, \textit{e.g.}, for a cluster's radius of gyration,
fractal dimension, or inertia tensor. Lastly, direct access to the
particles making up a cluster is provided.

As an example, let us consider the simulation of a ferrofluid
monolayer, i.e, a two-dimensional suspension of soft spheres carrying
a magnetic dipole at their center. The example is loosely based on
Ref.~\cite{cerda08a}. Magnetic particles tend to form chain and
ring-like clusters due to the non-isotropic nature of the
dipole-dipole potential.  For simplicity, we define particles as
\enquote{neighbors} if the distance between their centers is less
than $1.3\sigma$, where $\sigma$ denotes the particle diameter in the
purely repulsive Lennard-Jones potential~\cite{weeks71a}.  The sample
system contains 1000 magnetic particles at an area fraction of
$\phi=0.1$. The relative strength of the dipolar interactions to the
thermal energy is
\begin{equation}
\lambda=\frac{\mu_0 m^2}{4\pi \sigma^3 k_\mathrm{B} T},
\end{equation}
where $\mu_0$ denotes the vacuum permittivity and $m$ the particles'
dipole moment.  The dipolar interactions are calculated by means of
the dipolar P$^3$M method~\cite{cerda08d} and the dipolar layer
correction~\cite{brodka04a}.  While the particle positions are
confined to a plane, the magnetic dipoles can rotate freely in three
dimensions. This is called a quasi-2D system.

\begin{figure}[tb]
  \includegraphics[width=\linewidth]{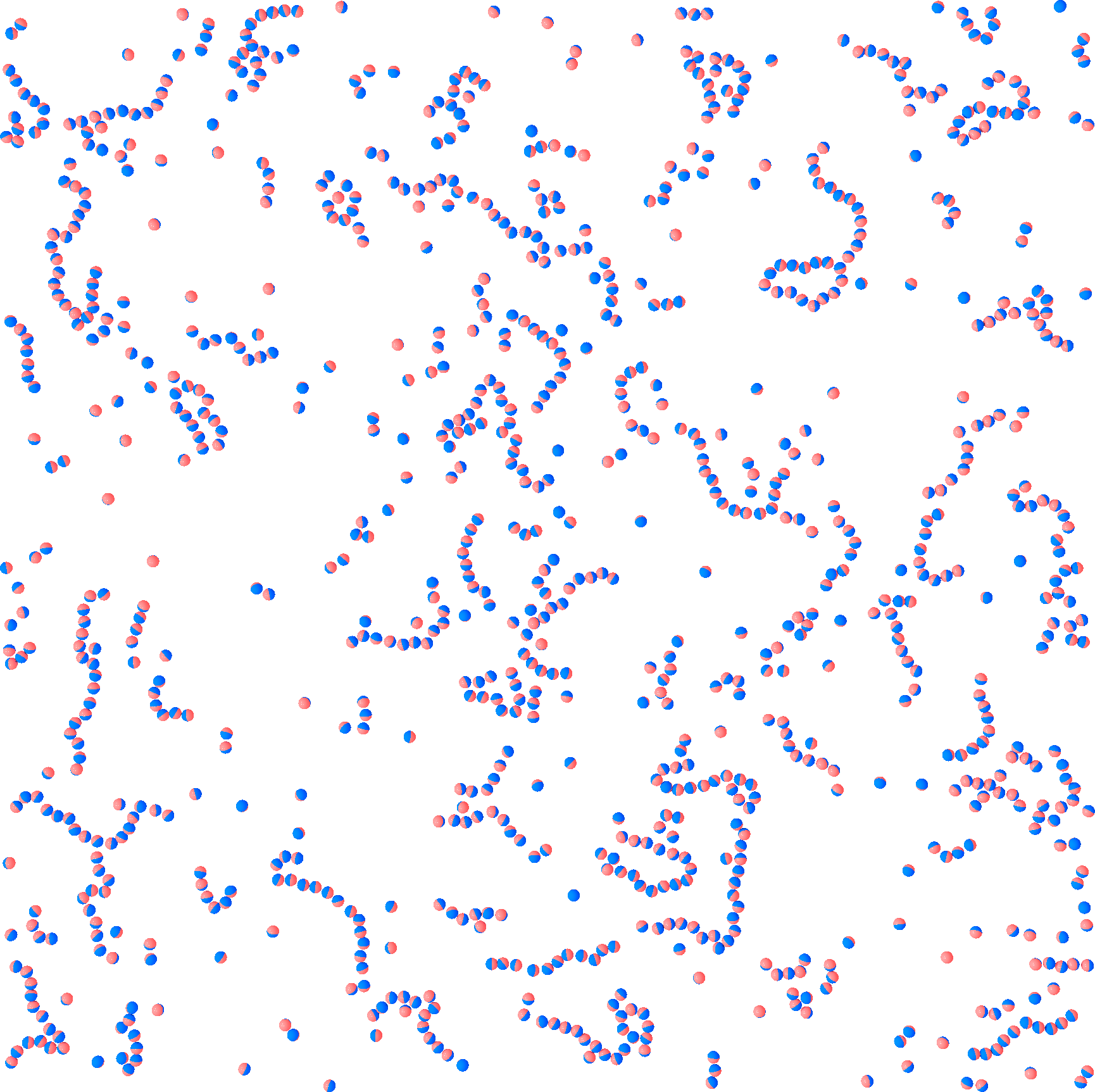}
  \includegraphics[angle=270,width=\linewidth]{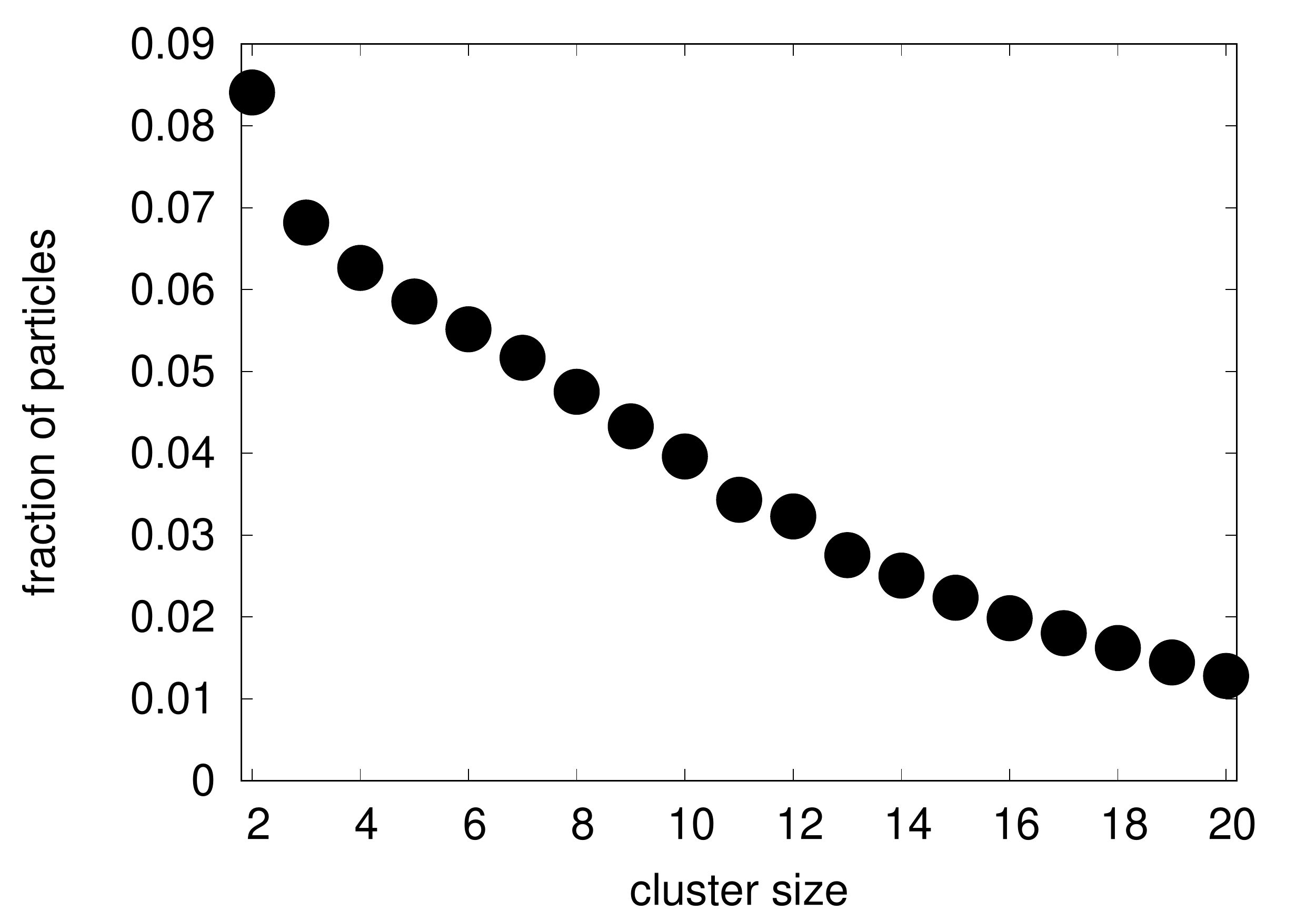}
  \caption{Top: Snapshot of the ferrofluid monolayer.  Bottom:
    Composition of the suspension plotted as fraction of particles
    being part of a cluster of a given size.}
  \label{fig:cluster}
\end{figure}

In Fig.~\ref{fig:cluster}, a snapshot from the simulation along with a
plot of the cluster composition of the suspension averaged over 1000
snapshots are shown. 
The full simulation script is provided in the
supplementary information. Please note that the script is meant as an
example. For a scientific study, larger systems, higher accuracies for
the dipolar interactions, and longer sampling times are needed.

\section{\label{sec:reaction}Particle-based Reactions}

Chemical reactions can take place in many systems, for example water
can autodissociate in positively charged protons and negatively
charged OH groups, or there are other groups that can dissociate only
partially in solvents like water, like the carboxylic group
COOH. These groups can be found, for example, in weak
polyelectrolytes, charge stabilized colloids, or
proteins~\cite{richtering06a}, and their charging state depends on the
pH-value~\cite{berg02a} of the solution. Therefore, the presence of
weak groups in charged polymers (polyelectrolytes) promote phenomena
like protonation-configuration coupling~\cite{castelnovo00a, shi12a}.
Protonation-configuration coupling means that the configuration of a
protein or polymer depends on the protonation state of the titratable
groups within the polymer. However, the protonation state itself
depends on the configuration. This tight coupling introduces the need
to investigate this phenomenon through computer simulations. Reactions
can also give rise to a net attraction of particles, which are on
average charge neutral~\cite{lund05a, lund13a}. Here, the net charge
of the particles fluctuates around its mean, thus allowing a particle
to be temporarily positive or negative. This fluctuation effect then
induces a net attraction. Investigating physical phenomena like the
above with the correct statistics, requires introducing particle based
reaction schemes in \esfour $\,$ which are described in the following.

\begin{figure}[tb]
    \centering 
    \includegraphics[width=\columnwidth]{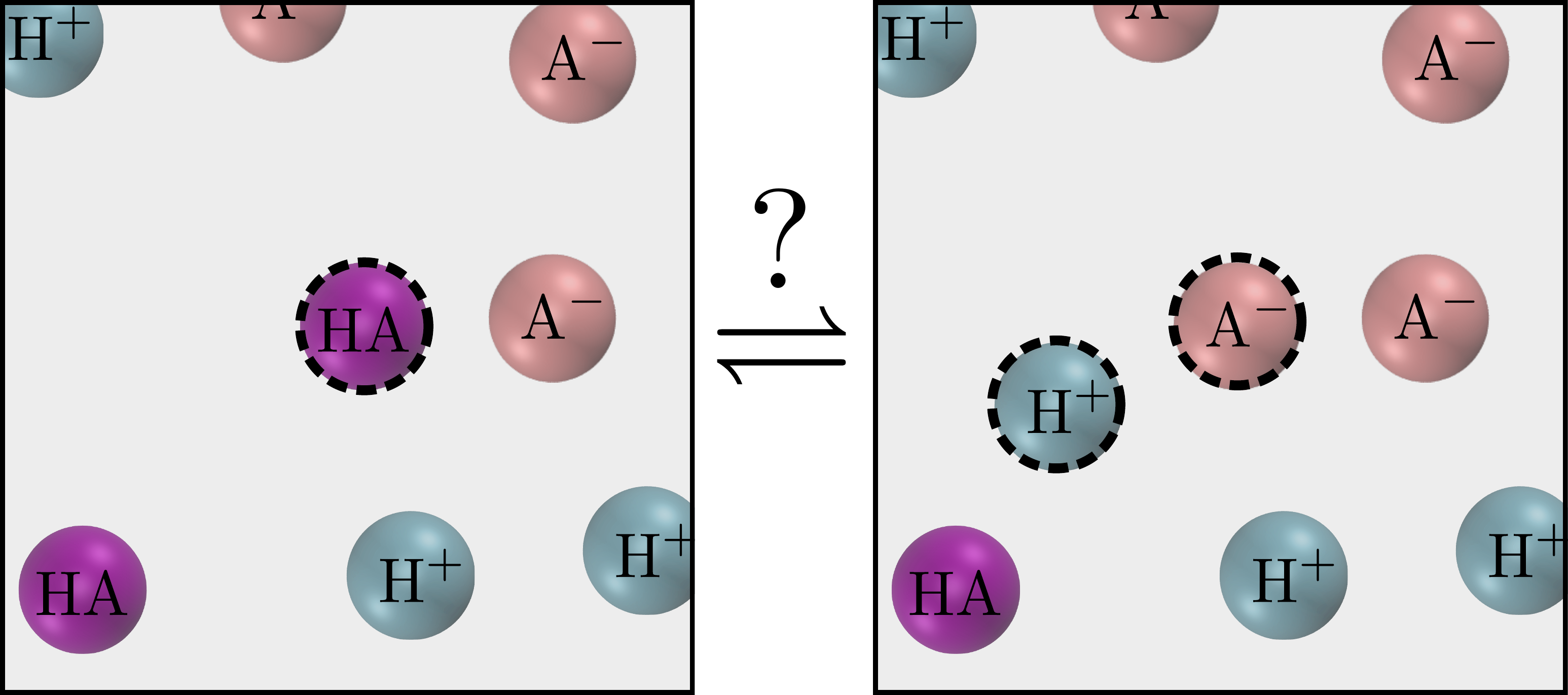} 
    \caption{The protonated/deprotonated states of a Monte-Carlo
      reaction step. Target particles are circled with a dotted line.}
    \label{fig:RxMC} 
\end{figure}

A typical acidic reaction is shown in Fig.~\ref{fig:RxMC}: an acid
particle ($\mathrm{HA}$) may release a proton ($\mathrm{H^+}$), or
vice versa, the deprotonated form of the acid ($\mathrm{A^-}$) may
take up a proton. Such protonation reactions change the electrostatic
charge of the particles taking part in the reaction. Therefore, \es{}
with its various electrostatic solvers~\cite{limbach06a}, is
well-suited to investigate electrostatic effects involved in
reactions.

The first scheme that allowed for acid-base reactions in thermodynamic
equilibrium using Molecular Dynamics and Monte Carlo simulations
appeared in the 1990s~\cite{reed92a} in the form of the constant-pH
ensemble. Soon after, the reaction ensemble~\cite{smith94a,
  johnson94a} was introduced which allows for the simulation of
arbitrary chemical reactions.  In both reaction schemes, particle
properties (such as the charge) need to be changed and
particles need to be created or deleted based on probabilistic
criteria. The reaction ensemble and the constant-pH method only
differ in the probabilistic criteria~\cite{landsgesell17b} and
both methods are implemented in \esfour.

In the reaction ensemble, arbitrary reactions can be implemented in the
simulation:
\begin{equation}
  \sum_{i=1}^z\nu_is_i=0,
\end{equation}
where $z$ chemical species of type $s_i$ with stoichiometric
coefficients $\nu_i$ are reacting~\cite{atkins10a}.  The acceptance
probability in the reaction ensemble for a reaction from state $r$ to
$l$ is given by~\cite{heath08a}
\begin{multline}
  \text{acc}^{\text{RE},\xi}(l|r) = \min\bigg\{1, (V\Gamma)^{\overline{\nu}\xi} \\
  \prod_{i=1}^z\left[ \frac{N_i^r!}{(N_i^r+\xi \nu_i)!}\right]\exp(-\beta \Delta E_{\text{pot},r \to l})\bigg\},
\end{multline}
where $N_i^r$ is the number of particles prior to a reaction and $\xi$
the \enquote{extent} of the reaction which is selected randomly with
$\xi \pm 1$ and $\beta=1/(k_\mathrm{B}T)$ proportional to the inverse
of the temperature. Further parameters are the concentration-dependent
reaction constant
$\Gamma=c^{\circ \overline{\nu}} \exp(-\beta \Delta G^\circ)$ where
$c^\circ$ is the reference concentration for which the change in the
free enthalpy $\Delta G^\circ$ is tabulated, the potential energy
difference with
$\Delta E_{\text{pot}, r\to l}=E_{\text{pot}, r}-E_{\text{pot}, l}$,
the volume of the system $V$ and the total change in the number of
molecules $\overline{\nu}=\sum_{i} \nu_i$ due to the reaction. The
corresponding protonation and deprotonation reactions are usually
performed after a fixed number of MD simulation steps with constant
particle numbers~\cite{heath08a}.

In addition to the above standard reaction ensemble
algorithm, we implemented a rare event sampling method on top of the
reaction ensemble~\cite{landsgesell17a}, called the Wang-Landau \cite{landau01a}
reaction ensemble algorithm. The Wang-Landau scheme is a so-called
flat histogramme technique that samples all states with equal probability. This modification of the algorithm might prove useful when
investigating systems with metastable states and rare transitions
between them.

In the other chemical equilibrium sampling method which is present in
literature---the constant-pH ensemble---the acceptance probability for a reaction $\ce{HA <=> A- + H+}$ is
given by:
\begin{multline}
  \text{acc}^{\text{cpH},\xi}(l|r) =\min \bigl(1,\exp \bigl[-\beta ( \Delta E_\text{pot} \\
  \pm (\ln(10)/\beta) \left(\text{pH}-\text{pK}_{a}\right))\bigr]\bigr),
\end{multline}
where $\mathrm{pH}$ is the pH of an implicitly imposed proton
reservoir, $\text{pK}_{a}=-\log_{10}{\exp(-\beta \Delta G^\circ)}$ and
where $\pm$ is used in the case of a
association/dissociation. Additionally the dissociation proposal
probability is proportional to the number of dissociable groups
$\mathrm{HA}$, while the association proposal probability is
proportional to the number of deprotonated groups $\mathrm{A^-}$.

\begin{figure}[tb]
    \centering 
    \includegraphics[width=\columnwidth]{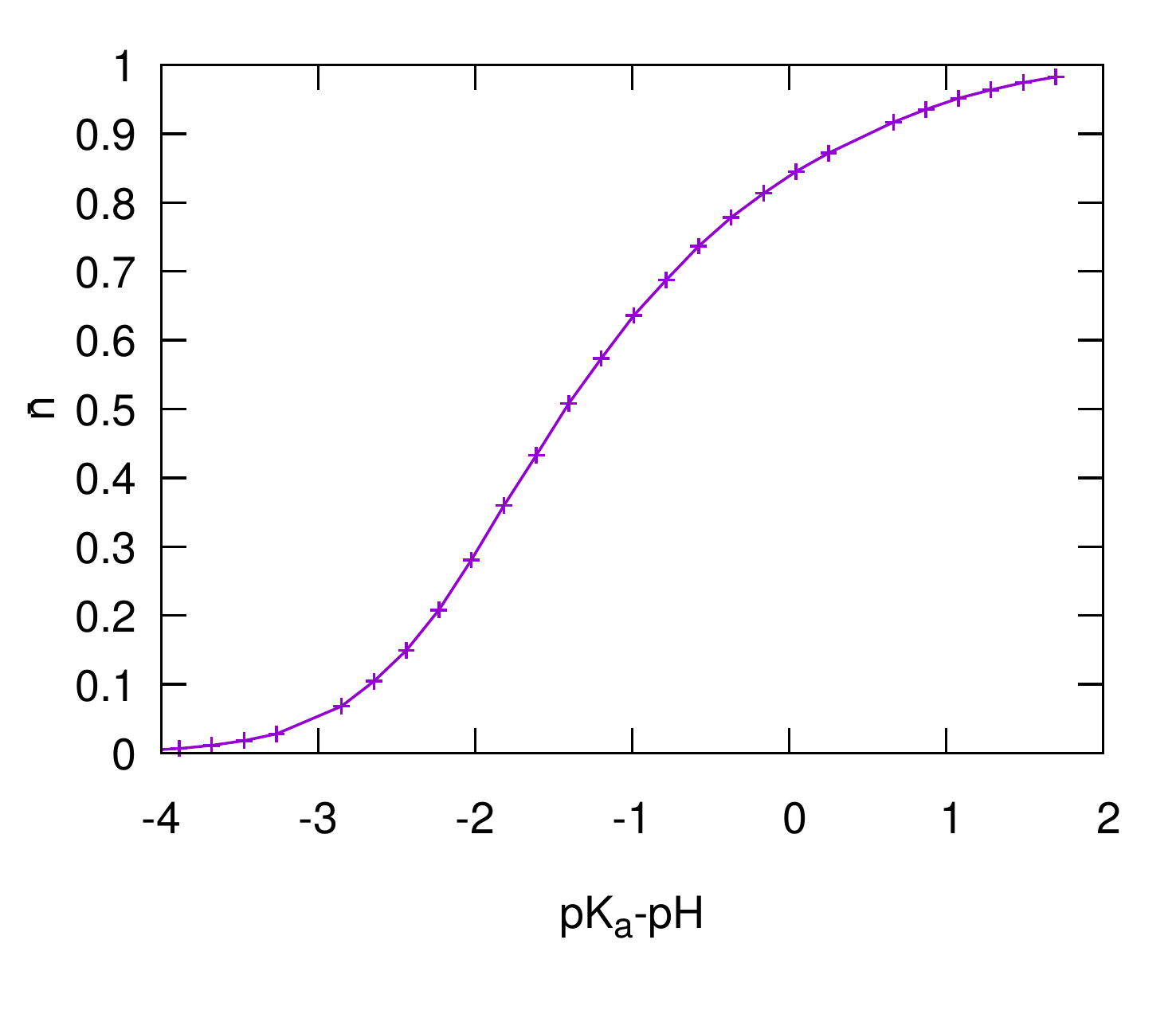} 
    \caption{The titration curve of a linear weak polyacid obtained with the constant pH method. For increasing pH the acid is less associated. The data is taken from Ref.~\cite{landsgesell17b} and simulation parameters are described there in more detail.}
    \label{fig:constant pH titration curve} 
\end{figure}

As a showcase for our ability to simulate chemcial reactions in \esfour \, we present in figure \ref{fig:constant pH titration curve} the titration curve of a linear weak polyelectrolyte.
A typical titration curve which is obtained from experiment shows the degree of dissociation as a function of $pK_a-pH$. The degree of association $\overline{n}$ measures how many particles $N$ are in the associated state $\mathrm{HA}$ or in the dissociated state $\mathrm{A^-}$: $\overline{n}=\frac{N(\mathrm{HA})}{N_{\mathrm{HA}}+N_\mathrm{A^-}}$. At low pH, a weak acid is mostly associated ($\overline{n}\approx 1$) while at higher pH a weak acid becomes less and less associated ($\overline{n} \approx 0$).

The reaction ensemble implementation in ESPResSo also allows for
simulations in the grand-canonical ensemble. This is, because the
grand canonical simulation scheme~\cite{frenkel96a} can be represented
as a reaction \ce{$\emptyset$ <=> A} with
$\Gamma=c^b_{A} \exp(\beta \mu^{\text{ex}, b}_{A})$, where $c^b_{A}$
is the bulk concentration of the species $A$ and
$\mu^{\text{ex}, b}_{A}$ the excess chemical potential of the species
in the bulk.  The needed excess chemical potential can be obtained via
the Widoms insertion method~\cite{frenkel96a}. We also implemented
this method in \esfour $\,$. It resembles a reaction which is
constantly rejected while observing potential energy changes due to
the insertion of particles.  The excess chemical potential in a
homogeneous system is then given by~\cite{frenkel96a}
\begin{equation}
  \mu^\text{ex}=-k_\mathrm{B}T \ln \Bigl(\Bigl\langle e^{-\beta
      (E_\text{pot}(N+1)-E_\text{pot}(N))} \Bigr\rangle
  \Bigr) .
\end{equation}

\section{\label{sec:drude}Including Explicit Dipolar Polarization with Drude Oscillators}

\begin{figure}[tb]
    \centering 
    \includegraphics[width=\columnwidth]{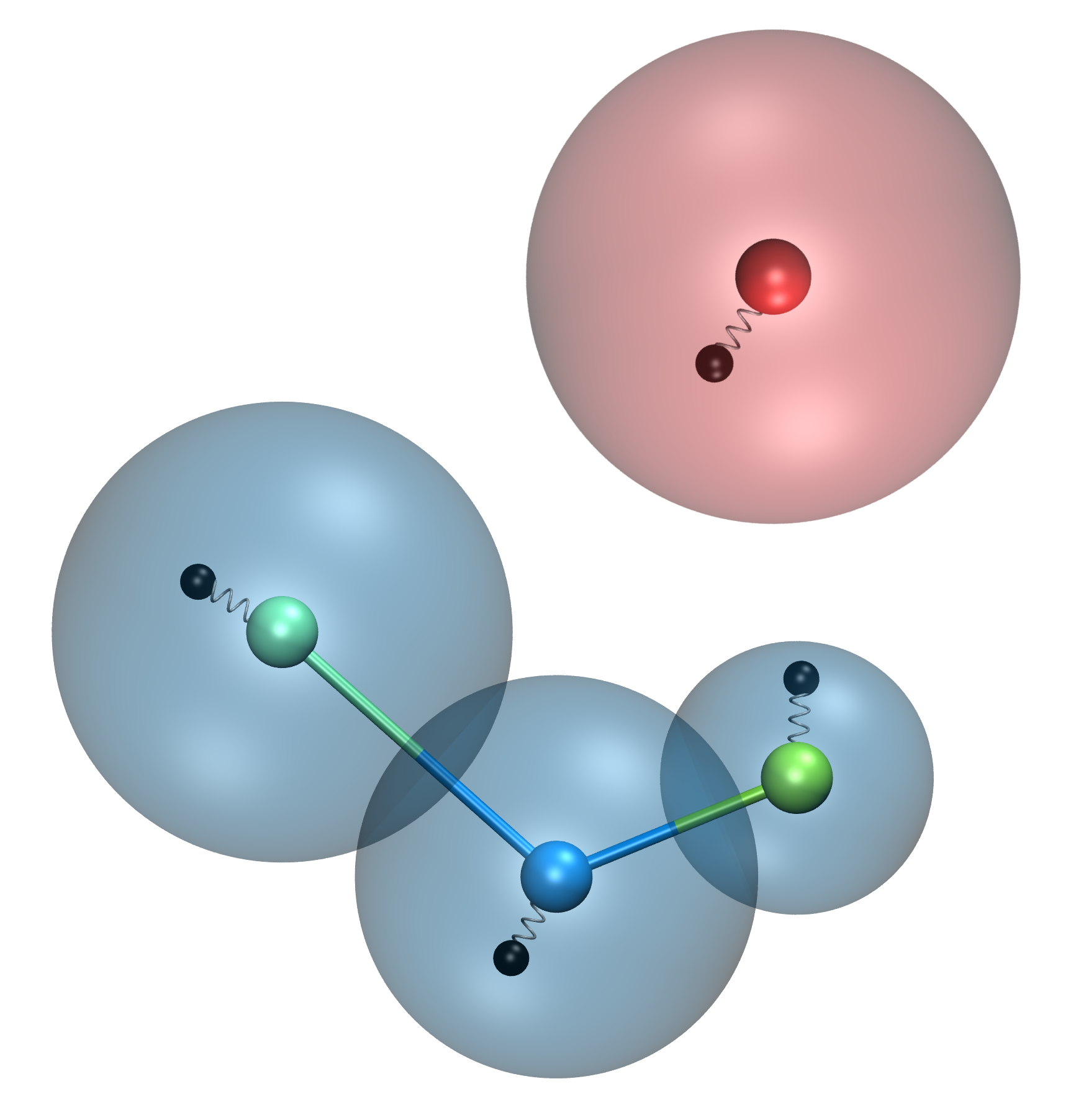} 
    \caption{Sketch of the polarizable coarse grained model for the
      ionic liquid BMIM PF$_6$.}
    \label{fig:bmimpf6_drude} 
\end{figure}
A particle's electron density is redistributed by the local electric
field, leading to a nonadditive molecular interaction. Especially the
study of ionic liquids (molten salts) polarization plays a large role
\cite{}. Often, this is
effectively included in the force-field (\textit{e.g.} via reduced
charges) to match continuum properties~\cite{leontyev11a,schmidt10a,dommert12a,dommert13a,kohagen16a}. In systems
where polarization effects are supposed to be inhomogeneous, e.g. at
interfaces, incorporating the dynamic nature of polarization can
refine the physical picture. Also simulations of bulk systems can
benefit from explicit polarization. For example, the experimental
values for the static dielectric constant and the self-diffusion
coefficient of water were accurately reproduced using a polarizable
model~\cite{lamoureux06b}.  Thermalized cold Drude
oscillators~\cite{lamoureux03a}, available in \esfour, can be used to
model this dynamic particle polarization. The basic idea is to add a
\enquote{charge-on-a-spring} (Drude charge) to a particle (Drude core)
that mimics an electron cloud which can be elongated to create a
dynamically inducible dipole (see Fig.~\ref{fig:bmimpf6_drude})\cite{bordin16a}. The
energetic minimum of the Drude charge can be obtained
self-consistently, which requires several iterations of the system's
electrostatics solver and is usually considered computationally
expensive~\cite{yu03a}.  However, with thermalized cold Drude
oscillators, the distance between Drude charge and core is coupled to
a thermostat, so that it fluctuates around the self-consistent field
solution.  This thermostat is kept at a lower temperature compared to
the global temperature to minimize the heat flow into the system. A
second thermostat is applied on the center of mass of the Drude charge
and core system to maintain the global temperature. The downside of
this approach is that usually a smaller time step has to be used to
resolve the high frequency oscillations of the spring to get a stable
system~\cite{mitchell93a}. In \es{}, the basic ingredients to simulate
such a system are split into three bonds; their combined usage creates
polarizable compounds:
\begin{enumerate}
    \item A \emph{harmonic bond} between charge and core.  
    \item For the cold thermostat, a Langevin-type \emph{thermalized
        distance bond} is used for the Drude-Core distance.
    \item A \emph{subtract P$^3$M short-range bond} to cancel unwanted electrostatic
    interaction.
\end{enumerate}
The system-wide thermostat has to be applied to the center of mass and
not to the core particle directly. Therefore, the particles have to be
excluded from the global thermostat, which is possible in \es{} by
setting the temperature and friction coefficient of the Drude complex
to zero. It thus remains possible to use a global Langevin thermostat
for non-polarizable particles. As the Drude charge should not alter
the charge or mass of the Drude complex, both properties have to be
subtracted from the core when adding the Drude particle.  In the
following convention, we assume that the Drude charge is always
negative. It is calculated via the spring constant $k$ and
polarizability $\alpha$ (in units of inverse volume) with
$q_d = \sqrt{- k \alpha}$.  For polarizable molecules (\textit{i.e.},
connected particles, coarse grained models etc.)  with partial charges
on the molecule sites, the Drude charges will have electrostatic
interaction with other cores of the molecule. Often, this is unwanted,
as it might be already part of the force-field (via partial charges or
parametrization of the covalent bonds). Without any further measures,
the elongation of the Drude particles will be greatly affected by the
partial charges of the molecule that are in close proximity. To
prevent this, one has to cancel the interaction of the Drude charge
$q_d$ with the partial charges of the cores $q_{\text{partial}}$
within the molecule. This can be done with the \emph{subtracts P$^3$M
  short-range bond}, which is also used to cancel the electrostatics
between Drude core $\leftrightarrow q_{\text{core}}$ and Drude charge
$q_d$.  This ensures that only the dipolar interaction inside the
molecule remains. The error of this approximation increases with the
share of the long-range part of the electrostatic interaction. In most
cases, this error is negligible comparing the distance of the charges
and the real-space cutoff of the P$^3$M electrostatics solver. In \es,
helper methods assist setting up this exclusion. In combination with
particle polarizability, the \emph{Thole correction}~\cite{thole81a}
is often used to correct for overestimation of induced dipoles at
short distances. Ultimately, it alters the short-range electrostatics
of P$^3$M to result in a damped Coulomb interaction potential
\begin{equation}
    V(r) = \frac{q_i q_j}{r} \biggl[1- e^{-s r} \biggl(1 + \frac{s r}{2}\biggr)\biggr].
    \label{eq:thole_potential}
\end{equation}
The Thole scaling coefficient $s$ is related to the polarizabilities
$\alpha_k$ and Thole damping parameters $a_k$ of the interacting
species via
\begin{equation}
    s = \frac{(a_i + a_j) / 2 }{ (\alpha_i \alpha_j)^{1/6} }.
    \label{eq:thole_scaling}
\end{equation}
Note that for the Drude oscillators, the Thole correction should be
applied only for the dipole part $\pm q_d$ added by the Drude charge
and not on the total core charge, which can be different for
polarizable ions.  Also note that the Thole correction acts between
all dipoles, intra- and intermolecular.  Again, the accuracy is
related to the P$^3$M accuracy and the split between short-range and
long-range electrostatics interaction. {\es} assists with the
bookkeeping of mixed scaling coefficients and has convenient methods
to set up all necessary Thole interactions.

\section{\label{sec:active}Active Particles}

\begin{figure}[tb]
\centering
  \includegraphics[width=\columnwidth]{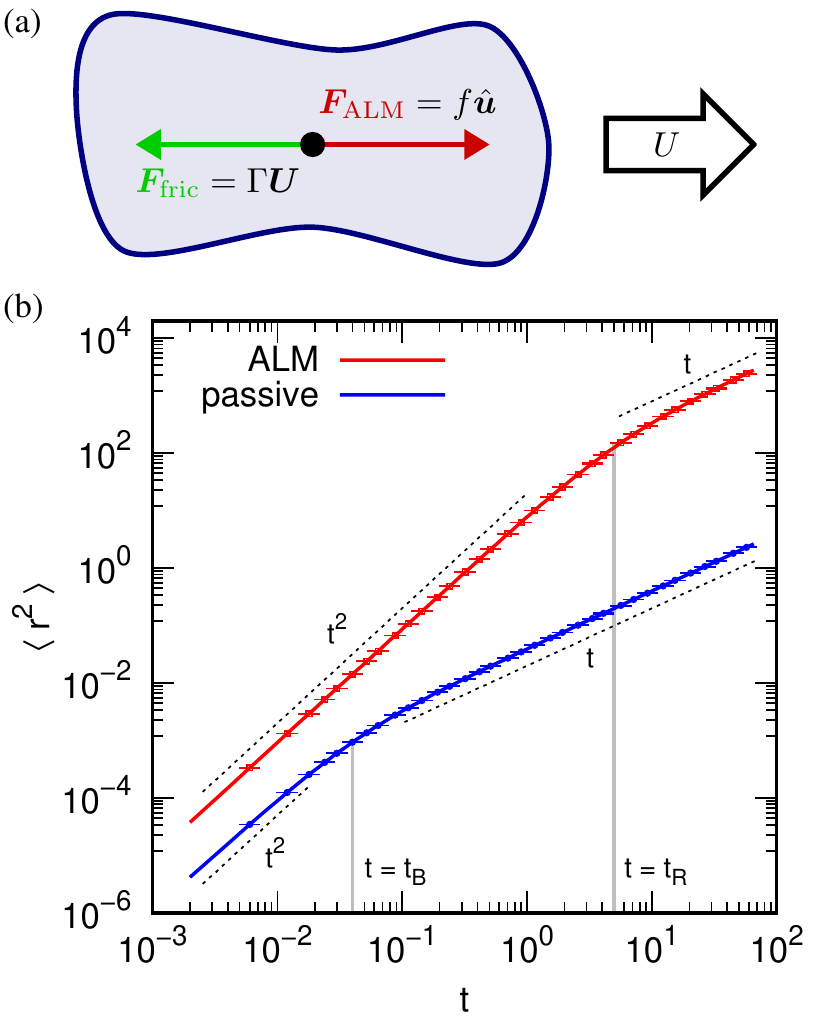}
  \caption{The Active Langevin Model (ALM). (a) A two-dimensional (2D) schematic of a shape anisotropic ALM self-propelled particle. The balance of self-propulsion force $f \hat{\bm{u}}$ and fluid friction $\underline{\bm{\Gamma}}_{t} \bm{U}$, leads to a constant swim speed $U$. (b) The mean-squared displacement $\langle r^{2} \rangle$ as a function of time $t$ of a passive Langevin particle (blue) and ALM swimmer (red), with error bars indicating the standard error. In the passive case, the ballistic regime $\langle r^{2} \rangle \propto t^{2} $ transitions to a diffusive regime $\langle r^{2} \rangle \propto t$ around the ballistic-to-diffusion crossover time $t_{B}$, as indicated by the grey vertical line. The ballistic regime is stretched by the activity and enhanced diffusion sets on a time scale associated with rotational diffusion $t_{R}$.}
  \label{fig:ALM}
\end{figure}

In \esfour{} it is possible to simulate active systems, wherein individual particles constantly transduce (internal) energy to perform work in the form of motion. The combination of this self-propulsion with particle interactions gives rise to unique out-of-equilibrium behaviors, of which living systems offer a wealth of examples: mammals, fish and birds~\cite{helbing00a, ballerini08a, katz11a, zhang13a, sliverberg13a}, as well as microorganisms~\cite{woolley03a, riedel05a, sokolov07b, polin09a, geyer13a, ma14b, reufer14a, schwarz16a}. The last decade and a half have seen the rapid development of man-made counterparts to microorganisms, following the first experimental realizations of \enquote{chemical swimmers} by~\citet{paxton04a} and~\citet{howse07a}, respectively. These artificial swimmers reduce the complexity of living matter, but still display signature features of being out of equilibrium, such as the ability to perform work~\cite{maggi16a} and motility-induced phase separation~\cite{theurkauff12a, palacci13a}.

Significant progress has been made both theoretically and computationally in understanding the individual and collective behavior of swimmers using simple models. Arguably the most famous of these is the Vicsek model~\cite{vicsek95a}, followed closely by the active Brownian model~\cite{ebeling99a, stenhammar13a, zheng13a}. In our developments, we have drawn inspiration from the Active Brownian Model (ABM) and created a variant thereof: The Active Langevin Model (ALM), which we will describe here. This ALM can also be coupled to the \es{} (GPU) lattice Boltzmann (LB) fluid dynamics solver~\cite{arnold13a, rohm12a} to account for the characteristic dipolar flow field that is associated with self-propulsion by microorganisms~\cite{drescher10a, drescher11a} as well as chemical swimmers~\cite{campbell18a-pre}. Our implementation~\cite{degraaf16a} is similar to the sub-lattice approach introduced by Nash~\textit{et al.}~\cite{nash08a, nash10a}. We refer to this extension as the hydrodynamic ALM (HALM) and we will briefly touch upon it here.

The ALM equation of motion for the translation of the $i$th swimmer's center of mass $\bm{r}_{i}$ is given by:
\begin{multline}
  \label{eq:EOMt}
  M \frac{\partial^{2}}{\partial t^{2}} \bm{r}_{i} = - \underline{\bm{\Gamma}}_{t} \frac{\partial}{\partial t} \bm{r}_{i} + f \hat{\bm{u}}_{i} \\
  - \sum_{j \ne i} \bm{\nabla} V(r_{ij},\bm{Q}_{i},\bm{Q}_{j}) + \bm{\xi}_{i,t}(t) .
\end{multline}
Here, we assume that the swimmer is fully shape anisotropic and we have introduced the following quantities: the mass $M$, the translational diffusion tensor $\underline{\bm{\Gamma}}_{t}$, a (co-rotating) unit vector that gives the direction of self-propelled motion $\hat{\bm{u}}_{i}$, the self-propulsion force $f$, and swimmer-swimmer interaction potential $V$. The latter depends on the separation $r_{ij} \equiv \vert \bm{r}_{i} - \bm{r}_{j} \vert$ and orientation of the swimmers, as specified by quaterions $\bm{Q}_{i}$. Thermal noise is introduced via $\bm{\xi}_{i,t}(t)$, which satisfies $\langle \bm{\xi}_{i,t}(t) \rangle = \bm{0}$ and $\langle \bm{\xi}_{i,t}(t) \otimes \bm{\xi}_{j,t}(t') \rangle = 6 k_{\mathrm{B}}T \underline{\bm{\Gamma}}_{t} \delta_{ij} \delta(t - t')$. The equation of motion for quaternions is lengthy and well-described in Ref.~\cite{martys99a}, it is therefore not reproduced here.

Figure~\ref{fig:ALM}a shows a representation of an ALM swimmer. Each particle is assigned a direction of self-propulsion $\hat{\bm{u}}_{i}$, along which it experiences a constant force $\bm{F}_{\mathrm{ALM}} = f \hat{\bm{u}}$. This self-propulsion force balances against the friction exerted by the implicit solvent, $\bm{F}_{\mathrm{fric}} = \underline{\bm{\Gamma}}_{t} \bm{U}$, leading to persistent directed motion with swim speed $U = \vert \bm{U} \vert$. Individual swimmers can have different propulsion forces and/or friction matrices in \esfour{}, providing users with tremendous flexibility in creating mixtures with a range of mobilities. 

A swimmer's direction of self-propulsion may be changed by rotational Brownian motion or by torques, either from external fields or collisions. The former leads to enhanced diffusion~\cite{howse07a} as shown in Fig.~\ref{fig:ALM}b. ALM in \esfour{} currently does not allow for the coupling of translational and rotational degrees of freedom, as required for the simulation of L-shaped~\cite{kuemmel13a} and chiral swimmers~\cite{wensink13a}; future releases will bring such functionality. Lastly, note that ALM is not fully damped (Fig.~\ref{fig:ALM}b),~\textit{i.e.}, there is an inertial component to the dynamics, which is uncommon in simulating (large) colloidal particles. However, ALM approaches the fully damped ABM when the friction coefficient is chosen to be large. In this limit, the time step should be suitably small to ensure that the algorithm produces the correct physics.

\begin{figure}[tb]
\centering
  \includegraphics[width=\columnwidth]{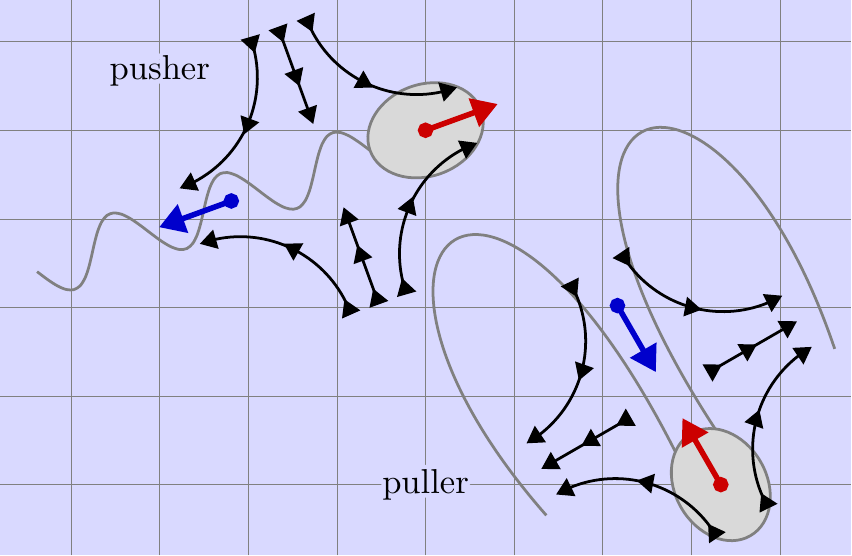}
  \caption{Sketches of two types of force-free hydrodynamic ALM (HALM)
    swimmers. Pusher and puller swimmers result from a change in the
    placement of the counter force (blue arrow) with respect to the
    swimmer's center, on which a propulsive force (red arrow) is
    applied. These forces are resolved sub-lattice (gray mesh) and
    interpolated to achieve the flow fields, as sketched using black
    arrows. The inherent friction of the swimmer's body with the
    lattice Boltzmann fluid leads to a persistent motion with a
    well-defined swim speed. The similarity of this figure to Fig.~4.1
    from R. Nash's PhD thesis~\cite{nash10a} is intentional and emphasizes
    the inspiration for our algorithm.}
  \label{fig:HALM}
\end{figure}

HALM introduces coupling between ALM and an LB fluid to account for hydrodynamic interactions between swimmers, particles, and obstacles. In developing HALM we have drawn inspiration from the work of Nash~\textit{et al.}~\cite{nash08a, nash10a}, who formulated a similar coupling to an LB fluid for the ABM. A directional self-propulsion force is applied to the particle, similar to the way this is done in ALM. In LB, a single point particle gains an effective radius~\cite{ahlrichs99a} or rather an inherent friction by coupling to the grid. This friction counter-balances the propulsive force leading to a constant swim speed. However, this also leads to force onto the LB fluid, while self-propulsion in a fluid should be force free. Therefore, a counter-force is applied to the LB fluid to ensure that momentum is not transferred into it globally, see Fig.~\ref{fig:HALM}. This force\slash counter-force pair induces a flow field with the typical leading-order dipolar contribution that characterizes self-propulsion. Figure~\ref{fig:HALM} illustrates two combinations of the force pair leading to the well-known puller and pusher dipolar flow fields, representing,~\textit{e.g.}, algae~\cite{drescher10a, harris13a} and spermatozoa~\cite{kantsler12a}, respectively. We refer to Ref.~\cite{degraaf16a} for full details of the implementation.

The reason for introducing ALM and HALM here, rather than relying on Brownian dynamics, is related to the way particles in \es{} couple to the LB fluid. Using these algorithms in tandem, the effect of the hydrodynamic interactions between swimmers may be disentangled, in much the same way as was done for passive particles~\cite{roehm14a}. Careful tuning of the relevant hydrodynamic parameters and the self-propulsion forces is key to reproducing low-Reynolds number hydrodynamic solutions, we refer to Ref.~\cite{degraaf17a} for an in-depth discussion. These authors have verified that the Nash~\textit{et al.} implementation~\cite{nash08a, nash10a} and HALM have the same near-field flow characteristics, long-range hydrodynamic retardation effects, and particle dynamics~\cite{degraaf17a-private}. Thus, the inertial component to HALM does not significantly affect the dynamics of our model.

ALM and HALM can both be used in conjunction with the raspberry method of creating shape-anisotropic particles~\cite{lobaskin04a, chatterji05a, fischer15a, degraaf15b}. This enables, for example, the study of the effect of polarization and roughness on motility-induced clustering~\cite{ilse16a}. HALM combined with raspberry particles leads to hydrodynamic multipole moments beyond the leading dipole moment for the self-propulsion of shape-anisotropic particles in fluids~\cite{degraaf16a}. This property has been exploited to determine the effect of these hydrodynamic moments on the motion of these swimmers in confining geometries~\cite{degraaf16b}, as well as their interaction with passive particles in their surrounding~\cite{degraaf16a, degraaf17a}. The use of raspberry particles in combination with HALM removes most of the lattice artifacts that point-particle HALM suffers from~\cite{degraaf16a}. We therefore recommend employing this combination within \es{} to properly account for any rigid body dynamics in a fluid, including rotational coupling due to shape anisotropies.


\section{\label{sec:conclude}Conclusions}

In this work, we outlined the major revisions that \esfour{} has
undergone, and the benefits that the user will experience from these
changes.  These include the reimplementation of the user interface in
Python and an overhaul of the core architecture to modern software
development paradigms.  We have also reported new features and
presented them accompanied by specific use cases.  This enables the
simulation of systems at the forefront of soft matter research,
including active matter and catalytic reactions.

We foresee a bright future for \es{} as a software package and as a
research project.  Together with our collaborators, we will further
improve the documentation and usability of this platform.  In
addition, algorithmic improvements are planned, which include:
(i)~Adaptive grid refinement to the electrokinetics code.  (ii)~A
load-balancing scheme for efficient simulation of heterogeneous
systems.  (iii)~Lees-Edwards boundary conditions for rheological
measurements.  We cordially invite new users to try out \es{} as a
simulation tool for their research and participate in its continued
development.

The current status of the package and the latest tutorials and
documentation can be found on the project website
\url{http://espressomd.org}, or on GitHub
\url{https://github.com/espressomd/}.


\section*{Acknowledgments}

We would like to acknowledge more than 100 researchers who contributed
over the last fifteen years to the \es{} software, and whose names can
be found on our website \url{http://espressomd.org} or in the
\verb|AUTHORS| file distributed with \es{}.  We are also grateful to
our colleagues of the SFB~716 for numerous suggestions for extending
the capabilities of \es{} and in helping us to improve our software.
As part of a collaboration~\cite{smiljanic18a-pre}, Milena Smiljanic
contributed code to the cluster analysis framework, particularly to the
analysis routines on the single cluster level.

CH, KS, and FW gratefully acknowledge funding by the German Science
Foundation (DFG) through the collaborative research center SFB~716
within TP~C5, the SimTech cluster of excellence (EXC~310), and grants
\mbox{HO~1108/25-1}, \mbox{HO~1108/26-1}, \mbox{AR~593/7-1},
\mbox{HO~1108/28-1}.  MK, JdG, and CH thank the DFG for funding
through the SPP~1726 Microswimmers--From Single Particle Motion to
Collective Behavior. JdG further acknowledges funding from an NWO
Rubicon Grant (\#680501210) and a Marie Sk{\l}odowska-Curie Intra
European Fellowship (G.A.~No.~654916) within Horizon 2020. The
simulations were partially performed on the bwUniCluster funded by the
Ministry of Science, Research and Arts and the Universities of the
State of Baden-W\"urttemberg, Germany, within the framework program
bwHPC.

\section*{Author Contributions}

Supervision: CH; Funding Acquisition, CH; Resources, CH; Conceptualization: CH, RW, and FW.
Writing: All authors contributed to the preparation of the manuscript.
Software and Validation: FW, RW, and KS are core developers of \es{}. The main
code contributors of individual features discussed in this article are the
following. Test infrastructure: KS, MK. Visualization: KB, MK. Particle-based
reactions: JL. Drude oscillators: KB. External fields: FW. Steepest descent
energy minimization: FW. Cluster analysis: RW. Active particles: JdG, HM. H5MD
parallel output: KS. VTF output: DS. A full
list of code contributions can be found at \url{https://github.com/espressomd/espresso/graphs/contributors}.


\bibliography{icp,paper}

\end{document}